\documentclass[fleqn,usenatbib]{mnras}

\usepackage[T1]{fontenc}
\usepackage{ae,aecompl}

\usepackage{graphicx}	
\usepackage{amsmath}	
\usepackage{amssymb}	
\usepackage{enumerate}
\usepackage{newtxtext,newtxmath}
\newcommand{\minus}{\scalebox{0.75}[1.0]{$-$}}
\newcommand{\bce}[1]{\textcolor{black}{\rm #1}}
\newcommand{\rev}[1]{\textcolor{black}{\rm #1}}

\title[Dusty Tails]{On the likely magnesium-iron silicate dusty tails of catastrophically evaporating rocky planets}

\author[Campos Estrada et al.]{
Beatriz Campos Estrada$^{1,2,3,4}$\thanks{E-mail:beatriz.estrada@nbi.ku.dk}, James E. Owen$^{4}$, Marija R. Jankovic$^{5}$, Anna Wilson$^{4}$ and Christiane Helling$^{2,3}$ \\
\\
$^{1}$Centre for ExoLife Sciences, Niels Bohr Institute, {\O}ster Voldgade 5, 1350 Copenhagen, Denmark \\
$^{2}$Space Research Institute, Austrian Academy of Sciences, Schmiedlstrasse 6, A-8042 Graz, Austria \\
$^{3}$TU Graz, Fakultät für Mathematik, Physik und Geodäsie, Petersgasse 16, A-8010 Graz, Austria \\
$^{4}$Astrophysics Group, Imperial College London, Blackett Laboratory, Prince Consort Road, London SW7 2AZ, UK\\
$^{5}$Institute of Physics Belgrade, University of Belgrade, Pregrevica 118, 11080 Belgrade, Serbia\\
}

\date{Accepted XXX. Received YYY; in original form ZZZ}

\pubyear{2023}

\begin{document}
\label{firstpage}
\pagerange{\pageref{firstpage}--\pageref{lastpage}}
\maketitle

\begin{abstract}
Catastrophically evaporating rocky planets provide a unique opportunity to study the composition of small planets. The surface composition of these planets can be constrained via modelling their comet-like tails of dust. 
In this work, we present a new self-consistent model of the dusty tails: we physically model the trajectory of the dust grains after they have left the gaseous outflow, including an on-the-fly calculation of the dust cloud's optical depth. We model two catastrophically evaporating planets: KIC\,1255\,b and K2-22\,b. For both planets, we find the dust is likely composed of magnesium-iron silicates (olivine and pyroxene), consistent with an Earth-like composition. We constrain the initial dust grain sizes to be  $\sim$\,1.25\,-\,1.75\,$\mu$m and the average (dusty) planetary mass-loss rate to be $\sim$\,3$\,M_\oplus \mathrm{Gyr^{-1}}$. Our model shows the origin of the leading tail of dust of K2-22\,b is likely a combination of the geometry of the outflow and a low radiation pressure force to stellar gravitational force ratio. We find the optical depth of the dust cloud to be a factor of a few in the vicinity of the planet. Our composition constraint supports the recently suggested idea that the dusty outflows of these planets go through a greenhouse effect--nuclear winter cycle, which gives origin to the observed transit depth time variability. Magnesium-iron silicates have the necessary visible-to-infrared opacity ratio to give origin to this cycle in the high mass-loss state. 
\end{abstract}




\section{Introduction}
\begin{figure*}
\begin{minipage}{\linewidth}
\includegraphics[width=\linewidth]{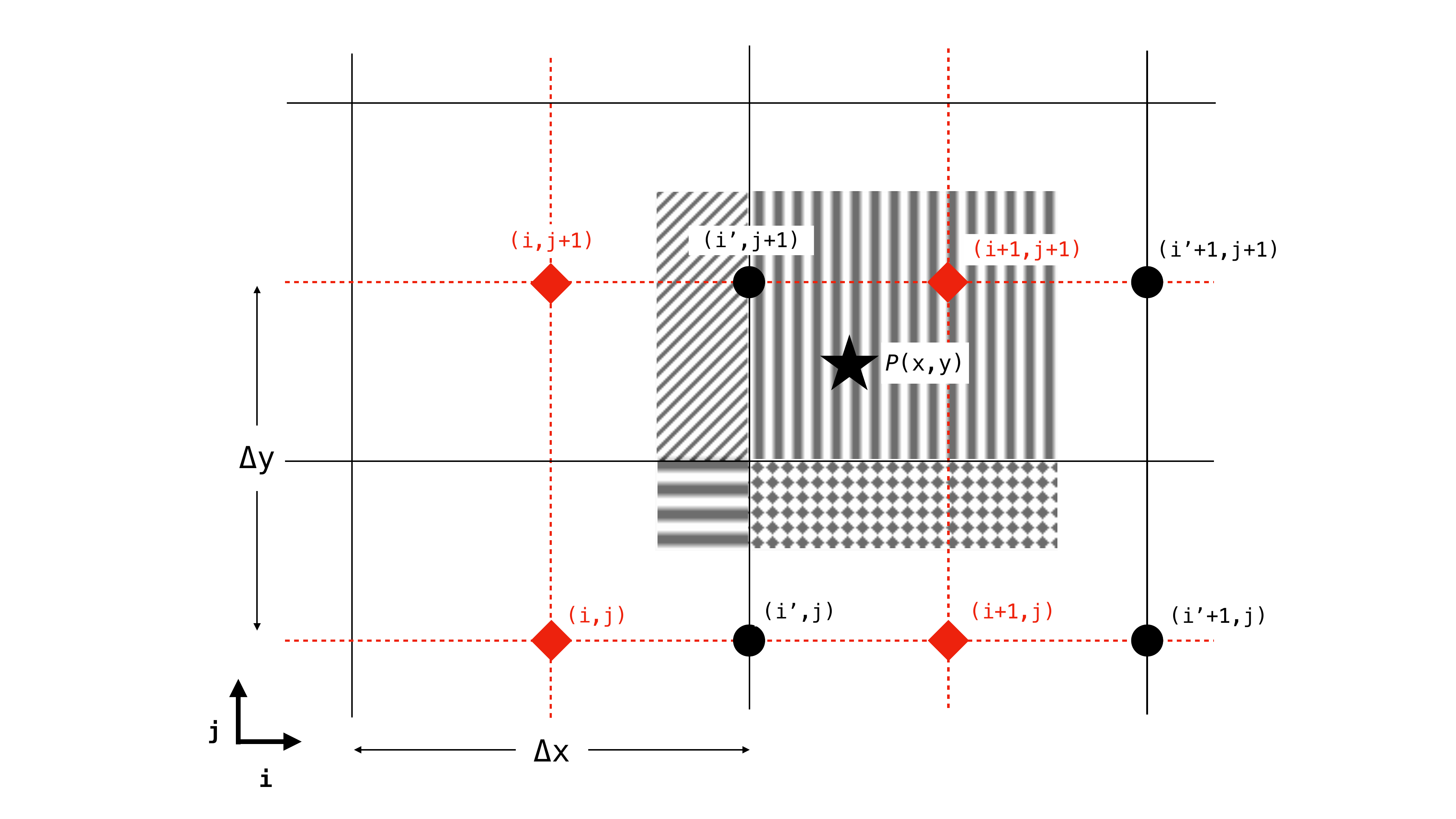}
\end{minipage}
 \caption{A 2D representation of our staggered mesh, similar to the 3D one used to estimate the dust cloud's optical depth. Scalar variables are measured at the centres of the grid cells, represented by the red diamonds. Vector quantities/direction-dependent variables are measured at the grid-cells interfaces, represented by the black circles. We apply the clouds-in-cells method (Section~\ref{sec:optical_depth}) where the extinction contribution of a dust grain in position $P(x,y)$ (black star) is shared between the grid-cells surrounding it (shaded area) with the value determined between the volume assigned to the grain and the volume of the cell.} 
 \label{fig:optical_depth_calculation}
\end{figure*}
The nature and composition of small planets' ($R_p\lesssim 1.5~$~R$_\oplus$) interiors remain uncertain. Measurements of planetary densities \citep[e.g.][]{Dressing2015, VanEylen2021} and modelling constraints \citep[e.g.][]{Rogers2015,Dorn2019,Gupta2019,Rogers2021,Rogers2023,Rogers2023b} show most  have an ``Earth-like'' composition of $2/3$ silicate rock and $1/3$ iron. However, density measurements alone do not provide enough information to determine a planet's bulk composition because planets with similar densities can vary greatly in terms of their bulk composition \citep[e.g.][]{Seager2007, Valencia2007, Rogers2010, Unterborn2016, Dorn2017,Dorn2017b,Neil2020,Neil2022}. One way around this degeneracy would be to constrain the elemental abundance ratios of the bodies. For example, this analysis can be performed for the material accreted by polluted white dwarfs \citep[e.g.][]{Gansicke2012,Farihi2016,Harrison2018,Bonsor2020}. However, it is uncertain what planetary reservoirs polluted white dwarfs probe \citep[e.g][]{Jura2014,Buchan2022,Brouwers2023}, and they are unlikely to represent close-in planets. 

Alternatively, ultra-short-period (USP) planets (planets with an orbital period of less than one day) provide a unique opportunity to study the composition of small planets. USP planets are tidally locked and highly irradiated by their host stars, often achieving sub-stellar (day-side) surface temperatures $\gtrsim 2000~$K. Their primordial atmosphere is expected to be lost quickly \citep[e.g.][]{Valencia2010,Owen2013}, leaving a ``bare'' core exposed to stellar radiation. Under these conditions, the surface of the planet becomes molten on its day-side, and a rock-vapour atmosphere forms from magma outgassing \citep[e.g.][]{Schaefer2009, Miguel2011, Ito2015, Kite2016, Mahapatra2017, Zilinskas2022}. Measuring the composition of this rock-vapour atmosphere could directly probe the planetary interior composition since the planet's atmosphere is likely in vapour pressure equilibrium with the magma. \citet{Zieba2022} recently argued the small ($1.5\,R_\oplus$) rocky USP planet K2-141\,b probably has a thin rock-vapour atmosphere. Further, \textit{JWST} observations will likely unveil more about the nature of these planets \citep[][]{Zilinskas2022}. However, current observations have yielded no evidence for thick atmospheres \citep{Kreidberg2019,Crossfield2022}. Furthermore, detailed modelling of these planets' interior--atmosphere interaction and evolution is required to interpret the observations and understand their planetary interiors.

In 2012, \citet{Rappaport2012} reported the discovery of KIC\,12557548\,b (often shortened to KIC\,1255\,b and more recently Kepler-1520\,b), a transiting object with an orbital period of approximately 15.7\,h which displays an asymmetric light curve. The KIC\,1255\,b light curve presents a sharp transit ingress and a slow egress. Additionally, the light curve shows a positive bump in stellar flux before the transit occurs and high transit depth time variability. \citet{Rappaport2012} suggested this light curve could be explained by the existence of a low-mass (similar to Mercury) evaporating USP planet with a comet-like tail of dust. The comet-like tail scenario can be explained as follows: dust from the planet's molten surface condenses in a thermally-driven wind \citep[][]{Rappaport2012,PerezBecker2013, Booth2023} and is transported by a gaseous outflow out to the point where the gas density is low enough that dust dynamically decouples from the gas. At this point, the trajectory of the dust is dictated by the stellar radiation pressure and the stellar gravity, which shapes the dust into a cometary tail, trailing the planet \citep[e.g][]{Rappaport2012, Brogi2012, Budaj2013, SanchisOjeda2015, vanLieshout2018}. As the gas density decreases away from the planet, the sublimation temperature decreases \citep[e.g.][]{Booth2023}. This results in the gradual sublimation of the dust as it moves along the tail, ultimately resulting in a tail where the optical depth drops with distance from the planet \citep{Rappaport2012, vanLieshout2016}.  
The optical depth pattern of the dust cloud can explain the sharp ingress and slow egress observed. Furthermore, the forward scattering of stellar photons by the dust grains towards the line-of-sight of the observer can give rise to the pre-transit brightening \citep[e.g.][]{vanLieshout2016}. This dusty-tail scenario has been validated by studies of the colour dependence of the transit depth \citep[e.g.][]{Bochinski2015}. In addition to this, \citet{vanWekhoven2014} obtained an upper limit on the KIC\,1255\,b's radius of 4600 km, supporting the idea that we are observing the dust cloud and not the planet. 

Since the discovery of KIC\,1255\,b, two more examples of evaporating rocky planets have been observed: KOI-2700\,b \citep{Rappaport2014} and K2-22\,b \citep[][]{SanchisOjeda2015}. KOI-2700\,b presents a very similar light curve to KIC\,1255\,b. K2-22b presents a symmetric light curve, with increased observed flux both before and after the transit, where the post-transit brightening is explained with a leading tail of dust \citep{SanchisOjeda2015}. 

To explain the high transit depth time variability, \citet{Rappaport2012} state an erratic variation in the dust production rate is needed, and a detailed understanding of the origin of this variability remains uncertain. \citet{PerezBecker2013} speculated and \citet{Booth2023} found the planetary outflow can be unsteady. For fast dust growth rates and moderate optical depths, there is a cycle between periods of dust production and no dust production, which could give rise to the observed transit depth variability. The variability was investigated further by \citet{Bromley2023}, following the speculation of heterogeneous condensation by \citet{Booth2023}. \citet{Bromley2023} demonstrated a chaotic evolution could arise when iron-poor silicates condensed at low-mass loss rates (low optical depths), while iron-rich silicate dust condensed at high mass-loss rates (high optical depths). 

Using models of the planetary outflow,  \citet[][]{PerezBecker2013}, \citet{ Kang2021} and \citet{ Booth2023} constrained the mass of KIC\,1255\,b and the planetary mass-loss rate. In addition to this, the models agree with a scenario where the outflows are marginally optically thick in the planet's vicinity. 
\begin{table*}
\caption{Dust species used in this study and the corresponding densities, sublimation parameters and references for optical data and sublimation parameters.}
 \begin{tabular}{l|l|l|c c c c |l}
     \hline
     \hline
     Dust species & Density  & Optical data & \multicolumn{4}{c|}{Sublimation parameters}  &Notes\\
     & [$\rm{g~cm^{-3}}$] & ref. &$\mathcal{A} ~\rm{[10^4~K]}$ & $\mathcal{B}$& $\alpha$ & ref.\\
     \hline
     $\rm{Al_{2}O_{3}[s]~(Corundum)}^{1}$ & 4.00 & K95& $\rm{7.74}$  & $\rm{39.3}$ & 0.1 & S04, L08 & \\
     $\rm{MgSiO_3}[s]~(Enstatite)^{1,2}$ & 3.20 &  J94, D95, J98 & $\rm{6.89}$  & $\rm{37.8 }$ & $0.1$ & M88 & (a)\\
     $\rm{Mg_{[0.5, 0.7, 0.95]}Fe_{[0.5,0.3,0.05]}SiO_3[s]}~(Pyroxene)^{1}$ & 3.20, 3.01, 2.74 &  J94, D95 & $\rm{6.89}$  & $\rm{37.8 }$ & $0.1$ & M88 & (a), (b)\\
     $\rm{Mg_{2}SiO_4}[s]~(Forsterite)^{1}$ & 3.27 & F01 & $6.53 $ & $34.1$ & $0.1$ & N94 & \\
     $\rm{Mg_{1.72}Fe_{0.21}SiO_4}[s]~(San~Carlos~Olivine)^{2}$ & 3.30 & F01, Z11 & $6.53 $ & $34.1$ & $0.1$ & N94 & (a), (c), (d) \\
     $\rm{Mg_{1.56}Fe_{0.40}Si_{0.91}O_4}[s]~(Sri~Lanka~Olivine)^{2}$ & 3.30 & Z11 &  $6.53 $ & $34.1$ & $0.1$ & N94 & (a), (c),(d) \\
     $\rm{Mg_{[1.0,0.8]}Fe_{[1.0,1.2]}SiO_4}[s]~(Olivine)^{1}$ & 3.71, 3.80 & J94, D95 &  $6.53 $ & $34.1$ & $0.1$ & N94 & (a), (d) \\
     $\rm{Fe_{2}SiO_4}[s]~(Fayalite)^{2}$ & 4.39 & F01 & $6.04$ & $37.7$ & $0.1$ & N94& (a) \\
     \hline
     \hline
 \end{tabular}
\label{tab:dust}
\end{table*}

\begin{table*}
       \begin{minipage}{\textwidth}
          {\textbf{Notes.} $^{1}$Amorphous. $^{2}$Crystalline. (a) Following \citet{vanLieshout2014}, $\alpha=0.1$ is adopted for materials with no evaporation coefficient measurement. (b) Following \citet{Kimura2002}, the sublimation parameters of enstatite are adopted for all other types of pyroxene. (c) The density chosen for this olivine is the average density of low-Fe olivine. (d) Following \citet{Kimura2002}, the sublimation parameters of forsterite are adopted for all other types of olivine. }
          \\
      \end{minipage}

      \begin{minipage}{\textwidth}
          {\textbf{References.} K95 \citet{Koike1995}; S04 \citet{Schaefer2004}; L08 \citet{Lirhmann2008}; J94 \citet{Jaeger1994}; D95 \citet{Dorschner1995}; J98 \citet{Jaeger1998};
          M88 \citet{Mysen1988}}; F01 \citet{Fabian2001}; N94 \citet{Nagahara1994}; Z11 \citet{Zeidler2011}. 
      \end{minipage}
\end{table*}
Morphological models of KIC\,1255\,b's dusty-tail reproduce the characteristics of the observed light curve and place constraints on the size of the dust grains by comparing synthetic light curves to the observed ones \citep[e.g.][]{Brogi2012, Budaj2013, vanWekhoven2014}{}{}. These models do not attempt to model the formation and launch of the dusty outflow but to study the properties of the dust under the assumption it is launched from the planets. The models indicate the dust grains are around micron-sized. 

K2-22\,b's dust tail properties have also been studied using similar morphological models. \citet{SanchisOjeda2015} determined K2-22\,b has a leading dust tail, giving origin to the observed post-transit brightening in flux. This implies the ratio of radiation pressure forces to stellar gravity ($\beta$) to be less than 2\%. The value of $\beta$ can be this small for low luminosity host stars and small particle sizes ($\lesssim 0.1 \mu$m) or large particle sizes ($\gtrsim 1.0 \mu$m). More recently, observations and models of K2-22\,b by \citet{Schlawin2021} show the average particle size must be larger than about 0.5\,-\,1.0$\mu$m, leading to mass-loss rates of about 1.6\,-\,1.8\,$M_\oplus\,\mathrm{Gyr^{-1}}$.
 
The dust in the planetary outflow is a direct sample of the magma pool on the planet's surface. Therefore, studying the dusty tails can be used to directly place constraints on the planetary composition. The composition of the dust was investigated in detail by \citet{vanLieshout2014, vanLieshout2016}. This work was done by assuming the planet launched an outflow with dust particles of a certain size and composition and studying how long they survived before they completely sublimated. \citet{vanLieshout2014} used the dust's survival timescale to compute the distance it could travel from the planet, and compared this to the length of the tail inferred from the light-curve. They found the models where the dust was composed of corundum ($\rm{Al_2O_3}$[s]) or iron-rich silicates to be consistent with the observations of KIC\,1255\,b and KOI-2700\,b. Pure iron, graphite or silicon carbide grains were ruled out. \citet{vanLieshout2016} directly simulated the dynamics and size evolution of dust particles in KIC\,1255\,b's tail to study the dust composition, the dust grain sizes and the planetary mass-loss rate of the planet. They found the dusty tail to be most likely composed of $\rm{Al_2O_3}$[s] micron-sized grains, with a planetary mass-loss rate ranging from 0.6 to 15.6 $M_{\oplus}\mathrm{Gyr^{-1}}$.  Enstatite ($\rm{MgSiO_3}$[s]) and forsterite ($\rm{Mg_2SiO_4}$[s]) had previously been suggested as potential compositions of the dust grains by \citet{Rappaport2012}, \citet{Brogi2012} and \citet{PerezBecker2013}. However, \citet{vanLieshout2016} found iron-free magnesium-silicates were too cold to sublimate efficiently due to their low optical, but high IR opacities, and hence survived too long in the dust cloud to be consistent with the observed transit duration.

All previous dusty-tail models are limited because they assume the dusty-tail to be optically thin throughout. In addition to this, many of the previous models use a simplified formulation for the complex refractive index in the calculation of dust optical properties. Considering the dust cloud to be optically thin throughout is an assumption that requires detailed investigation. The trajectory of a dust grain is highly dictated by the radiation pressure force, which is obviously highly sensitive to the optical depth to stellar light once it reaches moderate values. Thus, even moderate optical depths can cause significant changes in the morphology of the dusty tail. In addition to this, the temperature of a dust grain, and hence the sublimation rate of the dust, is highly sensitive to moderate or higher optical depths. Therefore, the more optically thick the environment, the cooler the dust grains will be, causing them to survive longer in the dusty tail.

In this work, we construct a new model of the dusty tails. We couple the dust particle dynamics to the self-consistently determined optical depth. Introducing the optical depth evolution allows for robust compositional constraints to be derived from the observations. Our methods are presented in Section~\ref{sec:model&methods}. In Section~\ref{sec:dust_comp} we apply our model to KIC\,1255\,b and in Section~\ref{sec:K222b} to K2\,22\,b.  In Section~\ref{sec:depth_variability}, we discuss our results in the context of the observations. Finally, in Section~\ref{sec:summary}, we summarise our findings.

\section{Model and methods}
\label{sec:model&methods}
\begin{figure}
\centering
\includegraphics[width=\linewidth]{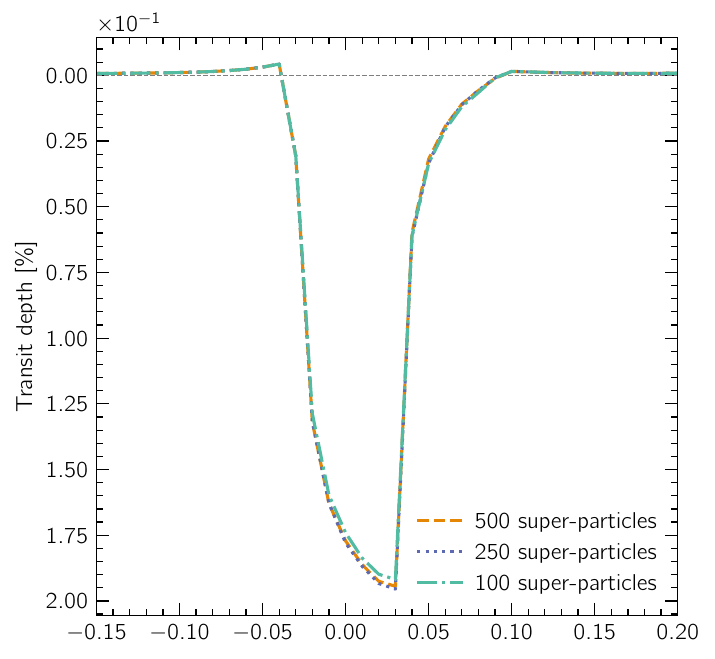}
\includegraphics[width=\linewidth]{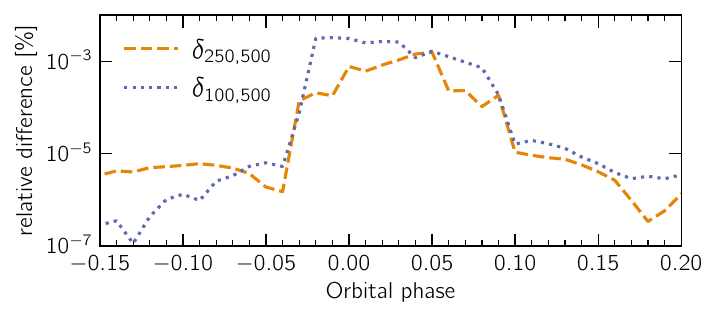}
 \caption{\textbf{Top:} Simulated transit of KIC\,1255\,b for different numbers of super-particles launched every $\rm{100^{th}}$ of an orbit. The dust is assumed to be composed of olivine ($\rm{Mg_{0.8}Fe_{1.2}SiO_4}$[s]). The dust cloud is assumed to be optically thin throughout ($\tau$=0.1) and the outflow is assumed to be radially outwards from the entire planetary surface (spherical outflow). \textbf{Bottom:} The relative percentage difference between the transits simulated on the top plot. The difference is smaller than the observed error ($\approx$ 200 ppm \citep{Rappaport2012}), therefore, we launch 250 super-particles every $\rm{100^{th}}$ of an orbit.}
 \label{fig:no_particles}
\end{figure}
We aim to develop a model with all the necessary physics to compare directly to the observations, but that is still computationally feasible. To achieve this, we adopt a hybrid method, where similar to \citet{vanLieshout2016}, the dust grains are modelled as Lagrangian ``super-particles'', but the optical depth is computed by ray-tracing on a grid of dust densities. Our super-particle is a collection of dust grains that share the same but evolving properties. The number of actual dust grains ($N_d$) inside a super-particle depends on the planetary mass-loss rate, the grain's initial size and composition, and how often/many super-particles we introduce into the simulation. Thus, we introduce the super-particle mass $m_{sp}=m_dN_d$, where $m_d$ is the mass of an individual dust grain.

\subsection{Dust motion}
The stellar radiation pressure force and the stellar gravity dominate the trajectory of the dust grains. For full consistency, we include planetary gravity and the Poynting-Robertson drag \citep{Roberston1937} acting on the dust grains, but we ignore collisions between the individual dust grains. Following the results from \citet{Booth2023}, we assume the planet to have a mass of $0.03M_\oplus$ although we note the choice does not affect our results. Using the \citet{Fortney2007} mass-radius relationship, we estimate the planetary radius to be $~0.33~R_\oplus$, assuming mass fractions of 2/3 silicate rock and 1/3 iron. The equation of motion for the dust grain is solved in the co-rotating frame of reference, centred at the centre of mass of the star-planet system. The planetary and stellar orbits are assumed to be circular. The equation of motion of a dust grain is given by:
\begin{equation*} 
    \ddot{\textrm{\textbf{r}}} = -\frac{G M_{\rm{\star}}}{{\mid \textrm{\textbf{r}} - \textrm{\textbf{r}}_{\star} \mid}^{ 3}} ( \textrm{\textbf{r}} - \textrm{\textbf{r}}_{\star}) -\frac{G M_{\rm{planet}}}{{\mid \textrm{\textbf{r}} - \textrm{\textbf{r}}_{\rm{planet}} \mid}^{ 3}} ( \textrm{\textbf{r}} - \textrm{\textbf{r}}_{\rm{planet}}) 
\end{equation*}
\begin{equation*}
\label{eq:motion}
   - \, \boldsymbol{\omega} \times (\boldsymbol{\omega} \times \textrm{\textbf{r}}) - 2(\boldsymbol{\omega} \times \dot{\textrm{\textbf{r}}}) 
\end{equation*}
\begin{equation}
\label{eq:motion}
+\, \beta \frac{ G M_{\rm{\star}}}{{{\mid \textrm{\textbf{r}} - \textrm{\textbf{r}}_{\star} \mid}^{ 2}}} \left [\left(1 - \frac{\dot{\textrm{\textbf{r}}}_{\rm{radial}}}{c} \right) \frac{( \textrm{\textbf{r}} - \textrm{\textbf{r}}_{\star})}{\mid \textrm{\textbf{r}} - \textrm{\textbf{r}}_{\star} \mid} - \left (\frac{\dot{\textrm{\textbf{r}}} + (\boldsymbol{\omega} \times \textrm{\textbf{r}})}{c} \right)\right]
\end{equation}
\noindent
where $\textrm{\textbf{r}}$ is the position vector of the dust grain, $\textbf{r}_{\star, \rm{planet}}$ are the position vectors of the star and planet, respectively; $M_{\star, \rm{planet}}$ are the mass of the star and planet respectively; $\boldsymbol{\omega}$ is the rotation vector of the frame-of-reference, $\beta$ is the ratio between the norms of the direct radiation pressure force and the star's gravitational force, $G$ is the gravitational constant and $c$ is the speed of light. The final term on the RHS of Equation~\ref{eq:motion} represents the radiation forces acting on the dust grains \citep[][]{Roberston1937, Burns1979}. 

We assume our dust grains are spherical, and as such, $\beta$ is given by
\begin{equation}
\label{eq:beta}
\beta = \frac{1}{4 \pi c G}\frac{L_{\star}}{M_{\star}}e^{-\tau_\star}\kappa(T_\star, s)
\end{equation}
\noindent where $L_\star$ is stellar luminosity, $\tau_\star$ is the optical depth of the dust cloud to stellar irradiation at the grain's position, and $\kappa$ is the radiation-pressure Planck-mean opacity at the stellar temperature $T_{\star}$, for a particle of radius $s$. 

\label{sec:dust_motion}
\subsection{Dust sublimation}
We take a similar approach to \citet{vanLieshout2016} for determining the sublimation rate of a spherical dust grain in a gas-free environment. 
In thermodynamical equilibrium, when the partial pressure of a molecule equals its equilibrium vapour pressure, the condensation rate must equal the sublimation rate; thus, the sublimation rate can be expressed in terms of the condensation rate using the principle of detailed balance \citep[c.f.][]{Booth2023}. In our models, the dust grains have large numbers of atoms ($\sim10^9$ atoms for a 0.1$\mu$m particle); thus, we assume their sublimation rate is solely dependent on the dust's internal properties (e.g. its temperature, \citealt{gail_sedlmayr_2013}). Consequently, the sublimation rate for a dust grain in thermodynamical equilibrium with the gas surrounding it is equal to its sublimation rate in a gas-free environment \citep{Langmuir1913}. Therefore, the change of the radius of a dust grain ($s$) can be written as \citep[][]{Kimura2002, gail_sedlmayr_2013, vanLieshout2014},
\begin{equation}
\label{eq:dust_sub}
\frac{{\rm D}s}{{\rm D}t} = -\frac{\alpha p_v(T_d)}{\rho_d} \left( \frac{\mu m_u}{2 \pi k_B T_d} \right)^{1/2}
\end{equation}
where $\alpha$ is the evaporation coefficient which parameterises the kinetic inhibition of the sublimation\footnote{$\alpha$ has been experimentally measured to be approximately the same as the growth/sticking coefficient \citep[][]{gail_sedlmayr_2013}{}{}.}, $p_v$ is the equilibrium vapour pressure, $\mu$ is the molecular weight of the sublimating molecules in atomic mass units\footnote{Following \citet{vanLieshout2014} and for simplicity, we assume $\mu$ to be the molecular weight of the dust sublimating. In reality, $\mu$ is the average molecular weight of the
molecules recondensing from the gas phase in the sublimation-condensation equilibrium.}, $m_u$ is the atomic mass unit, $k_B$ is Boltzmann's constant, and $T_d$ is the dust grain's temperature. The equilibrium vapour pressure is given by the Clausius-Claperyon relation \citep{Kimura2002},
\begin{equation}
p_v(T) = \exp(\mathcal{-A}/T + \mathcal{B})~\rm{[cgs]}
\label{eq:vapour_pressure}
\end{equation}
\noindent where $\mathcal{A}$ and $\mathcal{B}$ are material-specific  parameters. Quantities $\alpha$, $\mathcal{A}$ and $\mathcal{B}$ are assumed to be temperature independent. 

The balance between the received and emitted energy by a dust grain is used to determine the dust's temperature, which we approximate as:
\begin{equation}
\label{eq:dust_temp}
\kappa_{\rm{abs}}(T_\star,s)~ \frac{L_\star}{4 \pi r_d^2}~e^{-\tau_\star} = 4~\kappa_{\rm{abs}}(T_d,s)~\sigma_{SB}~T_d ^4
\end{equation}

\noindent where $\kappa_{\rm{abs}}(T_\star,s)$ is the Planck-mean absorption opacity at the stellar temperature $T_\star$, $\kappa_{\rm{abs}}(T_d,s)$ is the absorption opacity at the dust temperature $T_d$, $r_d$ is the dust grain-star distance and $\sigma_{SB}$ is Stefan-Boltzmann's constant. This equation assumes that while we are including the attenuation of stellar light due to the dust cloud's optical depth, we are assuming the dust cloud is optically thin to its own cooling radiation. This assumption is justified, as for our best fit models, the dusty outflow is optically thin to its re-emitted thermal emission. As discussed in our results, the dust's temperature (and hence sublimation rate) depends critically on the opacity ratio $\kappa_{\rm{abs}}(T_\star,s)/\kappa_{\rm{abs}}(T_d,s)$, which at higher ratios gives higher temperatures and hence faster sublimation. It is this opacity ratio that ultimately allows us to constrain the dust's composition and size through its impact on the temperature and, subsequently, the sublimation rate.

\label{sec:dust_sublimation}
\subsection{Optical depth evolution} 
\label{sec:optical_depth}
To calculate the optical depth of the dust cloud, we implement the clouds-in-cells (CIC) method \citep{cloudsincells}. Consider a staggered-mesh grid, where scalar variables (e.g. extinction) are defined at the grid-cell centres (red diamonds in Figure~\ref{fig:optical_depth_calculation}) and direction-dependent variables (e.g. optical depth) at the centres of the grid-cell interfaces (black circles in Figure~\ref{fig:optical_depth_calculation}). The extinction to be attributed to the grid-cell centres is obtained by sharing a dust super-particle's extinction over the grid points surrounding it. A 2D representation of our 3D method is shown in Figure~\ref{fig:optical_depth_calculation}. Considering a dust grain at a position $P(x,y)$, we can regard the grain as a small cloud and spread it over an area of $\Delta x \Delta y$, the shaded area surrounding the black star. The extinction in the horizontal-lines-shaded area is assigned to point $(i, j)$; that in the diamond-shaded area to point $(i+1, j)$; that in the vertical-lines-shaded area to point $(i+1, j+1)$ and that in the diagonal-lines-shaded area to point $(i, j+1)$.

We adapt and apply the method described above to a uniform three-dimensional spherical polar grid centred at the star, i.e. the volume of each grid cell is approximately the same. Here, $r$ is the radial distance from the star's centre, $\theta$ is the polar angle, and $\phi$ is the azimuthal angle. The contribution to the extinction of a super-particle at a position $(r_d, \theta_d, \phi_d)$ is given by
\begin{equation}
   \chi_{sp}(r_d, \theta_d, \phi_d) = \frac{ m_{sp} \, \kappa_{\mathrm{ext}}(T_{\star},s)}{\Delta V},
   \label{eq:chid}
\end{equation}
where $\kappa_{\mathrm{ext}}$ is the extinction opacity of the dust grains in a super-particle, and $\Delta V$ is the ``volume''\footnote{It's important not to confuse this ``volume'' with any physical volume. This ``volume'' is used to smooth the extinction over our grid in the CIC method and thus scales with the grid's resolution. Our choice for its size is confirmed by our resolution tests.} of the super-particle, taken to be the volume element of the cell in which it resides. For better numerical accuracy, volumes are calculated with the spherical volume element in the following form \citep[c.f.][]{Stone1992}:
\begin{equation}
\Delta V  = \Delta (R^3/3)~\Delta (-\cos\theta)~\Delta \phi. 
\end{equation}
The extinction of the grid cell centred at $(r(i),\theta(j),\phi(k))$ can now be written as
\begin{equation}
\chi (i, j, k) = \sum_{sp} V^f_{ijk} ~\chi_{sp}(r_d,\theta_d, \phi_d)
\label{eq:extinction}
\end{equation}
\noindent where $V^f_{ijk}$ is the fraction of the super-particle's volume that resides in cell $(i,j,k)$ (see Figure~\ref{fig:optical_depth_calculation}). The sum is performed over all super-particles that have any volume that overlaps with that cell.
Once the extinction grid is computed via Equation~\ref{eq:extinction}, we obtain the optical depth by numerically integrating the extinction over the radial distance from the star to the grid-cell interface centres in the radial direction $\textbf{i}$, using:
\begin{equation}
\tau (i,j, k) =\sum_{i' = 0}^{i-1}~\chi(i',j,k)~\Delta r,~\textrm{for } i \ge 1
\label{eq:tau_grid}
\end{equation} where $\tau (i,j,k)$ is the optical depth at the grid-cell interface centred at $(r(i),\theta(j), \phi(k))$ and $\Delta r$ is grid-cell size in the radial direction $\textbf{i}$. The grid boundaries are set as,
\begin{equation*} 
 r(i=0)= R_{\textrm{min}},~r(i=N_{\textrm{R}})= R_{\textrm{max}};
\end{equation*}
\begin{equation*} 
 \theta(j=0)= \theta_{\textrm{min}} + \frac{\Delta \theta}{2},~\theta(j=N_{\theta}-1)= \theta_{\textrm{max}} -  \frac{\Delta \theta}{2};
\end{equation*}
\begin{equation*} 
 \phi(j=0)= \phi_{\textrm{min}} + \frac{\Delta \phi}{2},~\phi(j=N_{\phi}-1)= \phi_{\textrm{max}} -  \frac{\Delta \phi}{2};
\end{equation*}
where $N_{\textrm{R}, \theta, \phi}$ are the number of grid-cells in the radial $\textbf{i}$, polar $\textbf{j}$ and azimuthal $\textbf{k}$ directions respectively, $(R, \theta,\phi)_{\textrm{min}}$ are the lower boundaries in each direction and similarly $(R, \theta,\phi)_{\textrm{max}}$ are the upper boundaries. We note that $r(i') = r(i) + \frac{\Delta r}{2}$.
In this case, at $i=0$, $\tau = 0$ for all $j$ and $k$. After testing, we set $N_{\textrm{R}} = 75$, $N_{\theta} = 25$ and $N_{\phi} = 250$, and the grid limits to be

\begin{equation*}
R_{\text{min}} = 0.95~a, ~R_{\text{max}} = 1.10~a;
\end{equation*}
\begin{equation*}
\theta_{\textrm{min}} = 1.55~\text{rad}, ~\theta_{\textrm{max}} = 1.60~\text{rad};
\end{equation*}
\begin{equation*}
\phi_{\textrm{min}} = \minus ~0.40~\text{rad}, ~\phi_{\textrm{max}} = 0.10~\text{rad};
\end{equation*}
where $a$ is the semi-major axis of the planet's orbit. These values allow for approximately cubic grid-cells with constant grid-cell sizes in each direction, where $\Delta r = 0.002\,a$ and $\Delta \theta = \Delta \phi = 0.002\,$rad. The grid limits were chosen according to where the optical depth is dynamically important. A lower limit of $0.001$ for the optical depth is placed. Outside the chosen grid limits, the optical depth of the dust cloud is set to 0.001. To obtain the optical depth at the position of a super-particle, we perform a tri-cubic spline interpolation \citep{numerical_recipes} over the optical depth grid.

\subsection{Dust opacities}
\label{sec:opacities}
Dust opacities are calculated for the materials listed in Table~\ref{tab:dust} as follows. First, using the \textsc{miescat} module of \textsc{radmc3dPy}\footnote{\raggedright\url{https://www.ita.uni-heidelberg.de/~dullemond/software/radmc-3d/manual_rmcpy/}}, which applies Mie theory, we calculate the dust absorption ($\kappa_{\rm abs}^\lambda$) and scattering opacities ($\kappa_{\rm sca}^\lambda$), as well as the scattering g-factor ($g_{\rm sca}^\lambda$), as functions of wavelength ($\lambda = 10^{-5}$--$10$\,cm) and grain size ($s=0.1$--$10$\,$\mu$m)\footnote{\raggedright{For some materials, in some wavelength ranges, optical constants are different for different crystal axes. In such cases, we follow \citet{vanLieshout2016} and combine the optical constants for different axes using the \citet{Bruggeman1935} mixing rule.}}. Where the optical constants are not available for the entire wavelength range indicated above, they are extrapolated by keeping them constant at shorter wavelengths and with a log-log function at longer wavelengths, as would be expected from simple diffraction theory \citep[e.g.][]{Bohren1983}. The extrapolation of optical constants to shorter wavelengths may somewhat affect the opacities at the stellar temperatures, especially for corundum ($\rm{Al_2O_3}$[s]), for which the optical constants are only available down to 0.5\,$\mu$m, but they do not change our conclusions. At the lower end of relevant temperatures, the calculated values are unaffected by the extrapolation since the available data covers the relevant range of wavelengths.

For each grain size $s$, the opacities are averaged over a Gaussian size distribution centred at $s$, and of width $\Delta \textrm{ln} (s)=0.02$ to avoid ``ringing''. We adopt the dust bulk densities shown in Table~\ref{tab:dust}.

To obtain the Planck-mean opacities as functions of temperature and grain size, we integrate $\kappa_{\mathrm{abs}}^\lambda$, $\kappa_{\mathrm{sca}}^\lambda$ and $g_{\mathrm{sca}}^\lambda$ over frequency. In our problem, we have four relevant opacities: 
    \begin{enumerate} 
    \setlength{\itemindent}{2em}
     \item The absorption opacity to stellar light, $\kappa_{\mathrm{abs}}(T_\star,s)$, used in Equation~\ref{eq:dust_temp};
    \item The absorption opacity at the grain's temperature, $ \kappa_{\mathrm{abs}}(T_d,s)$, used in Equation~\ref{eq:dust_temp};
    \item The extinction opacity to the star light, $\kappa_{\mathrm{ext}}(T_\star,s) = \kappa_{\mathrm{abs}}(T_\star,s)+ \kappa_{\mathrm{sca}}(T_\star,s)$, used in Equation~\ref{eq:chid};
    \item The radiation pressure opacity, $\kappa(T_\star,s) = \kappa_{\mathrm{abs}}(T_\star,s) + \kappa_{\mathrm{sca, eff}}(T_\star,s)$, used in Equation~\ref{eq:beta}, where $\kappa_{\mathrm{sca, eff}}(T_\star,s)$ is given by \ 
    $(1-g_{\mathrm{sca}}(T_\star,s)\,)\,\kappa_{\mathrm{sca}}(T_\star,s)$.
    
    \end{enumerate}

\subsection{Numerical methods}

\label{sec:numerical_method}

To determine the trajectory of a super-particle, we solve its equations of motion (Equation~\ref{eq:motion}) and sublimation (Equation~\ref{eq:dust_sub}) simultaneously.
To solve equations \ref{eq:motion} and \ref{eq:dust_sub}, we use the Dormand-Prince (DP) method \citep{Dormand1980} with an absolute and relative tolerance of $10^{-8}$. 
We apply Brent's method \citep{Brent1973} to solve Equation~\ref{eq:dust_temp} numerically for the temperature of a dust grain with an absolute tolerance of $10^{-4}$ and a relative tolerance of $10^{-8}$. Each super-particle has an optimal timestep for the DP solver. In order to make the model as efficient as possible, we use the optimal timestep for each individual super-particle and synchronise the timestep between particles when obtaining the optical depth of the dust cloud and calculating the transit depth (see Section~\ref{sec:light_curve}). This approach only updates the optical depth at every synchronised time step; however, we confirmed this does not affect our results by varying the synchronised time step. 

The model's free parameters are the distribution of the grain's initial positions and velocities, the initial dust grain sizes, the planetary mass-loss rate and the dust composition. 
To ensure the dust grains are not gravitationally bound to the planet, we start them off at a distance of 1.1 Hill radii from the planet's centre. We assume the dust grains have left the planetary atmosphere at the thermal velocity of the gas particles \citep{Booth2023} in the radial direction. The super-particles are either randomly distributed to leave from the entire planetary surface (spherical outflow) or just the planet's day-side, \bce{both at the level of the Hill radius (see Section~\ref{sec:K222b})}. 

The simulations are initialised by obtaining the optical depth grid of the initial dust cloud. The planetary mass-loss rate is set to be constant throughout our simulations. We launch 250 super-particles every 100$^{\rm{th}}$ of a planetary orbit, enough to guarantee convergence in the light curve (see Figure~\ref{fig:no_particles}). We trace the optical depth every $\rm{200^{th}}$ of an orbit. We consider a dust grain completely sublimated when it is smaller than $0.1\mu m$, at which point it does not significantly contribute to the extinction or scattering of the dust cloud.  We run the simulations until the transit profile is converged, i.e. it has reached a steady state and does not significantly change from one transit to the next; this typically requires five orbits (or until it becomes apparent there is no steady state for a given material; see Section~\ref{sec:MgSi}).

\subsection{Transit profile calculation}
\label{sec:light_curve}
\begin{figure*}
\centering
\begin{minipage}[b]{0.482\linewidth}
\includegraphics[width=\linewidth]{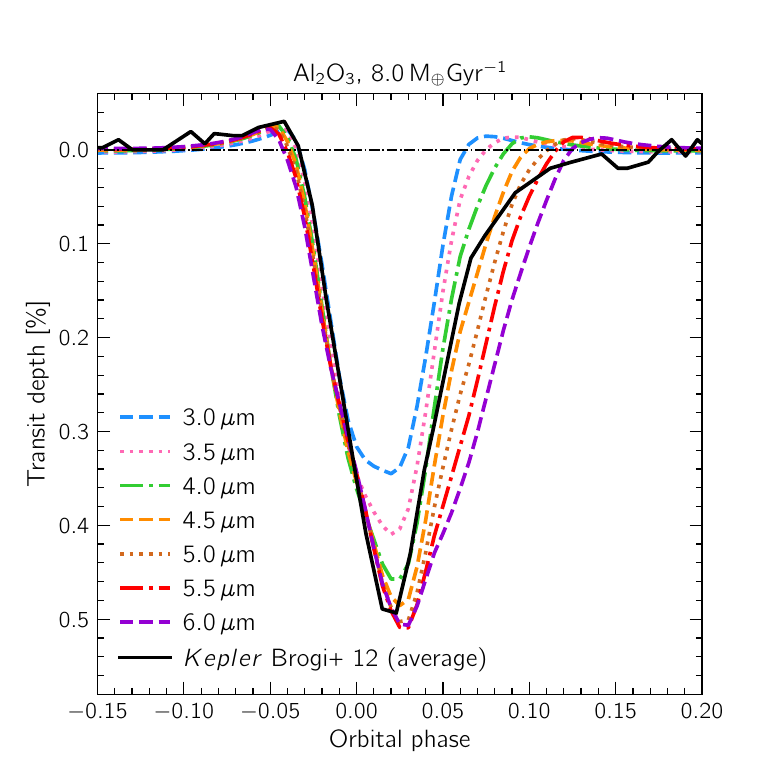}
\end{minipage}
\hspace{0.5cm}
\begin{minipage}[b]{0.482\linewidth}
\includegraphics[width=\linewidth]{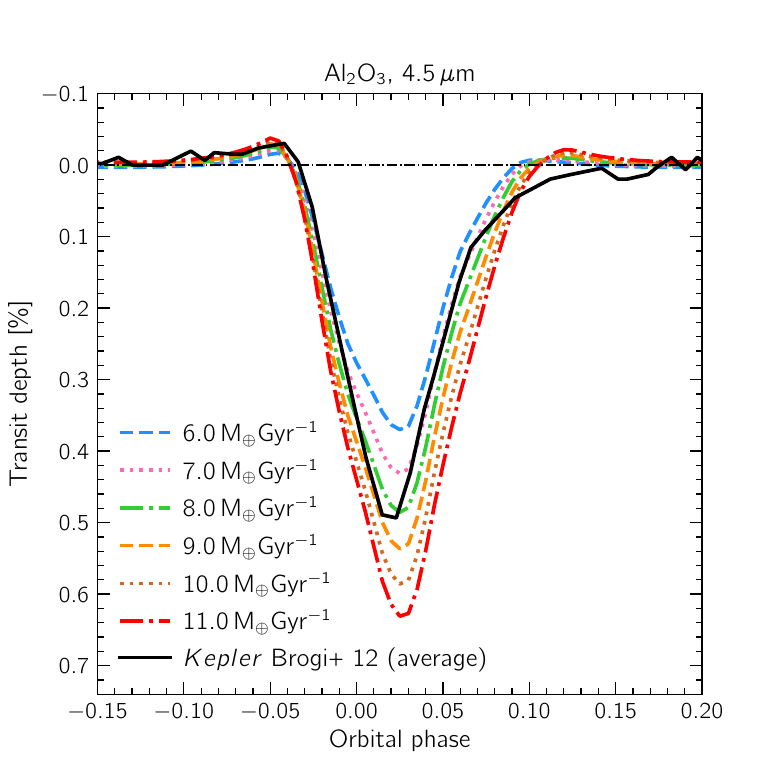}
\end{minipage}
 \caption{Synthetic light curves of KIC\,1255\,b for dust composed of corundum ($\mathrm{Al_2O_3}$[s]). The observed \textit{Kepler} light curve averaged over six quarters by \citet{Brogi2012} is shown as the black solid curve. \textbf{Left}: Synthetic light curves for models at a constant planetary mass-loss rate of 8.0$\,M_{\oplus}\mathrm{Gyr}^{-1}$ and different initial dust grain sizes. \textbf{Right}: Synthetic light curves for models at a constant initial dust grain size of 4.5$\,\mu$m and different planetary mass-loss rates.}
 \label{fig:Al2O3}
\end{figure*}
The final stage is to compute a synthetic light curve from our simulations. We adopt a two-stage method, including extinction and forward scattering. For the extinction calculation, we implement the CIC method once again (see Section~\ref{sec:optical_depth}). We grid the star into cells with area $A_{{\rm{cell}}}$ and spread each super-particle over an area $A_{{\rm{cell}}}$. If a super-particle overlaps with a given grid cell, the optical depth contribution from the super-particle in that cell is given by 

\begin{equation}
    \tau_{sp} = \frac{f\,\kappa_{\mathrm{ext}}(T_{\star}, s) \, m_{sp}}{A_{\rm{cell}}},
\end{equation}
 where $f$ is the fraction of the super-particle's area which overlaps with the grid cell. To obtain the total optical depth $\tau$ in a grid cell, we sum all the individual $\tau_{sp}$'s corresponding to that cell. \bce{The flux from that cell is then attenuated by a factor of $e^{-\tau}$.} Finally, the total stellar flux is obtained by adding all the individual fluxes from each grid cell. 

Forward scattering of light by dust grains can increase the observed stellar flux. The increase in flux is proportional to the scattering opacity of the dust grain, $\kappa_{\rm sca}(T_\star,s)$ (Section~\ref{sec:opacities}), and the scattering phase function at the scattering angle. We use the Henyey-Greenstein analytical scattering phase function for dust grain mixtures \citep{Henyey1941}, which is solely dependent on $g_{\rm{sca}}$ and the scattering angle, and use the single scattering approximation \citep{vanLieshout2016}. We ignored limb-darkening effects for both the extinction and forward scattering computations. The extinction and forward scattering components are combined into synthetic transit profiles.

\section{KIC 1255 b: Dust composition, grain sizes and mass-loss rate constraints}
\label{sec:dust_comp}
\begin{figure*}
\centering
\begin{minipage}[b]{0.48\linewidth}
\includegraphics[width=\linewidth]{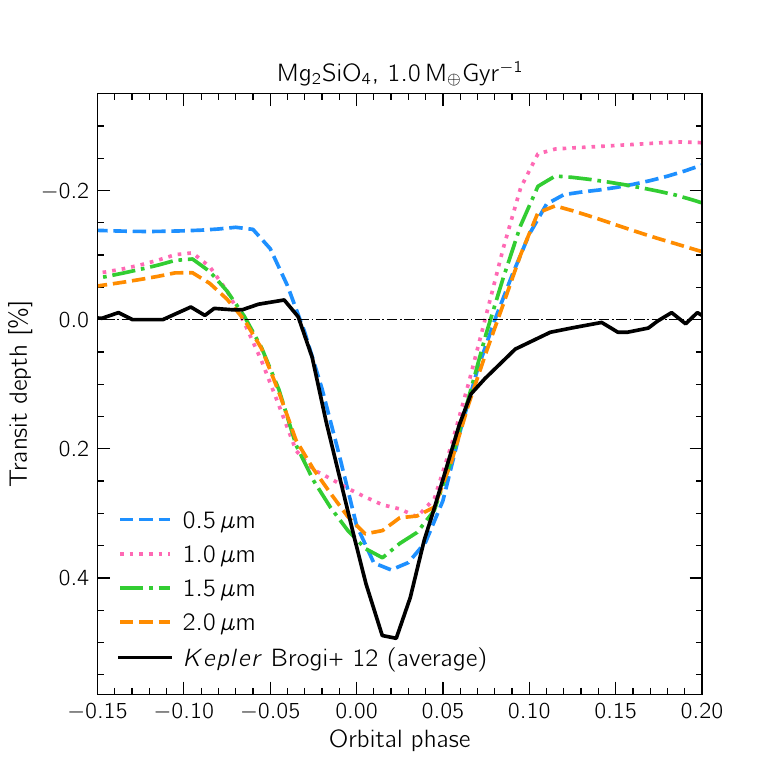}
\end{minipage}
\hspace{0.5cm}
\begin{minipage}[b]{0.48\linewidth}
\includegraphics[width=\linewidth]{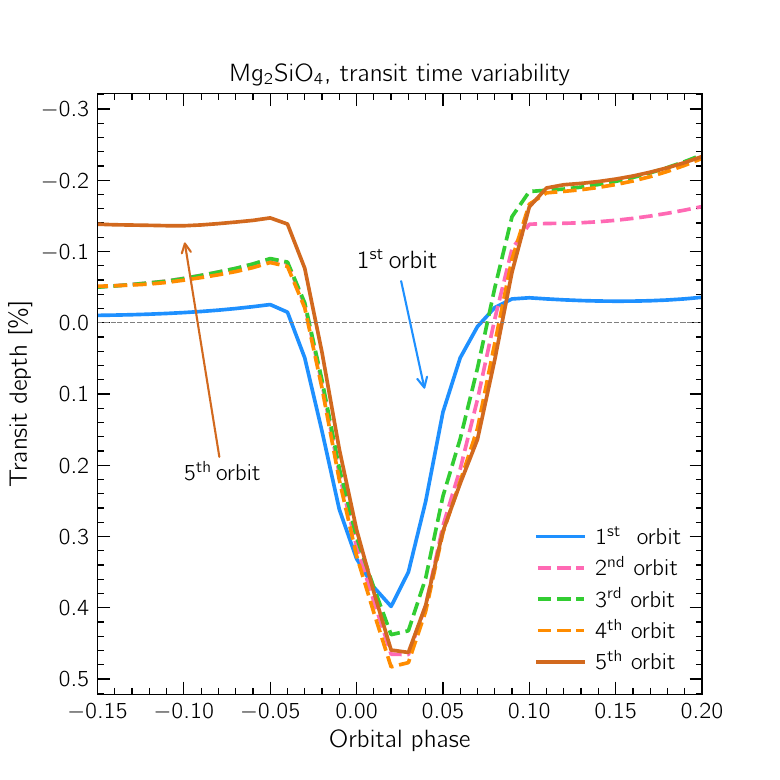}
\end{minipage}
 \caption{Synthetic light curves of KIC\,1255\,b for dust composed of forsterite ($\mathrm{Mg_2SiO_4}$[s]). \textbf{Left}: Synthetic light curves for models at a constant planetary mass-loss rate of 1.0$\,M_{\oplus}\mathrm{Gyr}^{-1}$ and different initial dust grain sizes after 5 orbits. The observed \textit{Kepler} light curve averaged over six quarters by \citet{Brogi2012} is shown as the black solid curve. \textbf{Right}: Synthetic light curves over 5 orbits at an initial dust grain size of 0.5$\,\mu$m and a planetary mass-loss rate of 1.0$\, M_{\oplus}\mathrm{Gyr}^{-1}$. Note how the light curve does not converge, particularly for the pre-transit brightening region.}
 \label{fig:forsterite}
\end{figure*}
\begin{figure*}
\centering
\begin{minipage}[b]{0.48\linewidth}
\includegraphics[width=\linewidth]{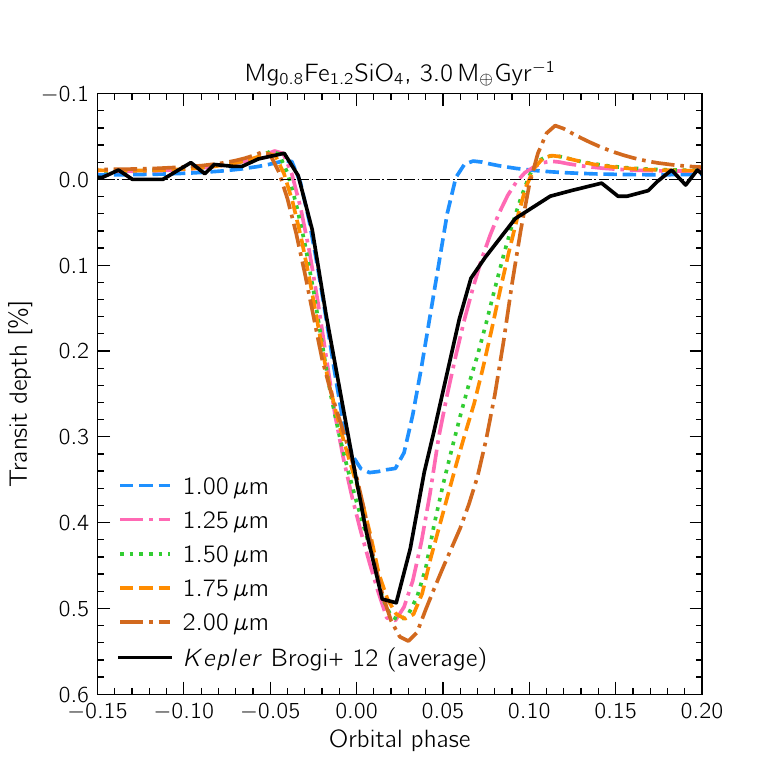}
\end{minipage}
\hspace{0.5cm}
\begin{minipage}[b]{0.48\linewidth}
\includegraphics[width=\linewidth]{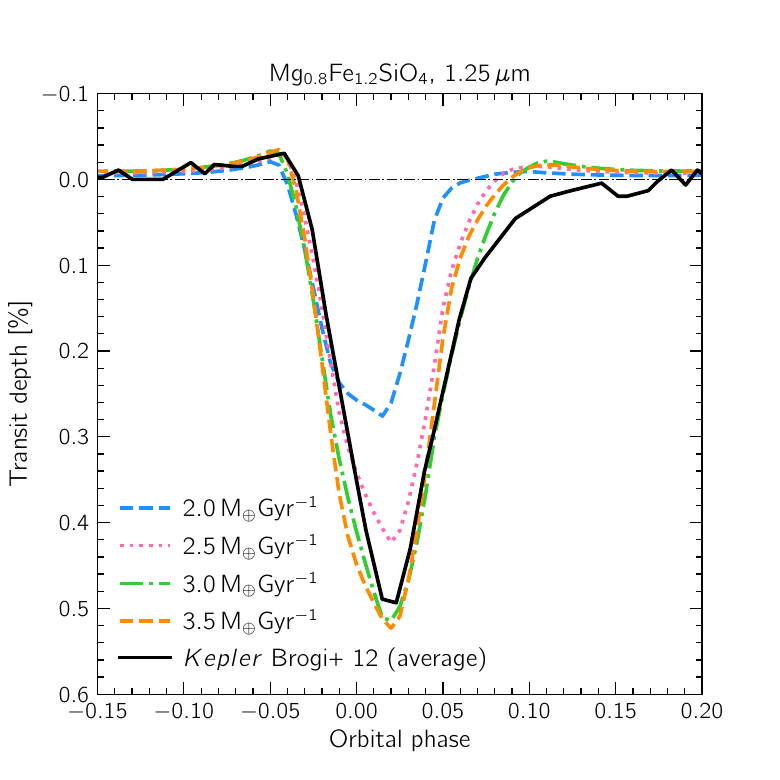}
\end{minipage}
 \caption{Synthetic light curves of KIC\,1255\,b for dust composed of iron-rich olivine ($\mathrm{Mg_{0.8}Fe_{1.2}SiO_4}$[s]). The observed \textit{Kepler} light curve averaged over six quarters by \citet{Brogi2012} is shown as the black solid curve. \textbf{Left}: Synthetic light curves for models at a constant planetary mass-loss rate of 3.0$\,M_{\oplus}\mathrm{Gyr}^{-1}$ and different initial dust grain sizes. \textbf{Right}: Synthetic light curves for models at a constant initial dust grain size of 1.25$\,\mu$m and different planetary mass-loss rates.}
 \label{fig:Mg08Fe12SiO4}
\end{figure*}
\bce{One of our key goals is to determine the dust composition and size. The size of the dust grains has been constrained through measurements of the transit depth dependence on wavelength, i.e. colour dependence. \citet{Croll2014} and \citet{Schlawin2016} observed the transit depths of KIC\,1255\,b not to be colour dependent and set a lower limit of the dust grain size of $\sim$0.2-0.5$\,\mu$m. Previous modelling efforts \citep[e.g.][]{Brogi2012, Budaj2013}{}{} predict the dust grain sizes to be 0.1\,-\,1.0\,$\mu$m based on the shape of the forward-scattering bump. However, these models did not consider the sublimation rate of the dust grains and simply assumed an exponential decay of the size distribution. Furthermore, \citet{vanLieshout2016} curiously found corundum ($\text{Al}_2 \text{O}_3\,$[s]) to be the best fit with the observations for KIC\,1255\,b, using a model that didn't include the dust tail's optical depth self-consistently on the dust dynamics.  }

\bce{Therefore, using our new, self-consistent approach, we constrain the tail's dust composition, grain sizes and planetary mass-loss rate. We run a grid of models and compare the synthetic light curves with observations}. \bce{The dust's optical properties, which are highly dependent on the dust grain sizes and compositions, control the morphology of the white light curve, specifically the pre-transit brightening and the long egress. On the other hand, the planetary mass-loss rate mainly influences the transit depth (see Section~\ref{sec:mdots}). We define the best-fit models to be the ones where the pre-transit brightening and the long egress shapes are best represented with a transit depth comparable to that of the average observed light curves.}
\bce{Therefore, the reported mass-loss rate is an attempt to estimate the \textit{average} dusty mass-loss rates (see Section~\ref{sec:mdots}).}

\rev{Since the planet itself is not detected, we do not know exactly when, in its orbit, the planet is at mid-transit in the observations. Thus, one needs to align the models with the observations by defining when an orbital phase of 0 occurs. We do this by aligning our synthetic light curves with the observations by the pre-transit brightening (if necessary). Additionally, we also align our out-of-eclipse flux to that of the observations.}

\bce{In our simulations, we compute the transit depth every 5 minutes. To obtain the synthetic light curves to compare to the observations, we smooth the data over the \textit{Kepler} long cadence (30 minutes)}. 

We first apply our model to KIC\,1255\,b. KIC\,1255\,b orbits a K-type star with \,$T_{\star}\approx 4550 \, K$, $R_{\star}\approx0.66\,R_{\odot}$\, and $M_{\star}\approx0.67\,M_{\odot}$ \citep[][]{Thompson2018}{}{}.
We run the model for the dust compositions listed in Table~\ref{tab:dust}. We study the parameter space of initial dust grain sizes ranging from 0.5 to 8.0$\,\mu$m, and (dusty) planetary mass-loss rates from 1.0 to 15.0$\,M_\oplus \text{Gyr}^{\minus 1}$. The models we discuss throughout Section~\ref{sec:dust_comp} assume the outflow is spherical. We also explored models with a day-side outflow geometry, but we found the spherical outflow to consistently be a better fit for KIC\,1255\,b \bce{(Appendix~\ref{ap:dayside}). This is because the day-side outflow models produce a more symmetric transit, which is less consistent with  KIC 1255 b's light curve}. We discuss the issue of outflow geometry further in Section~\ref{sec:K222b}. 

\begin{figure*}
\centering
\begin{minipage}[b]{0.48\linewidth}
\includegraphics[width=\linewidth]{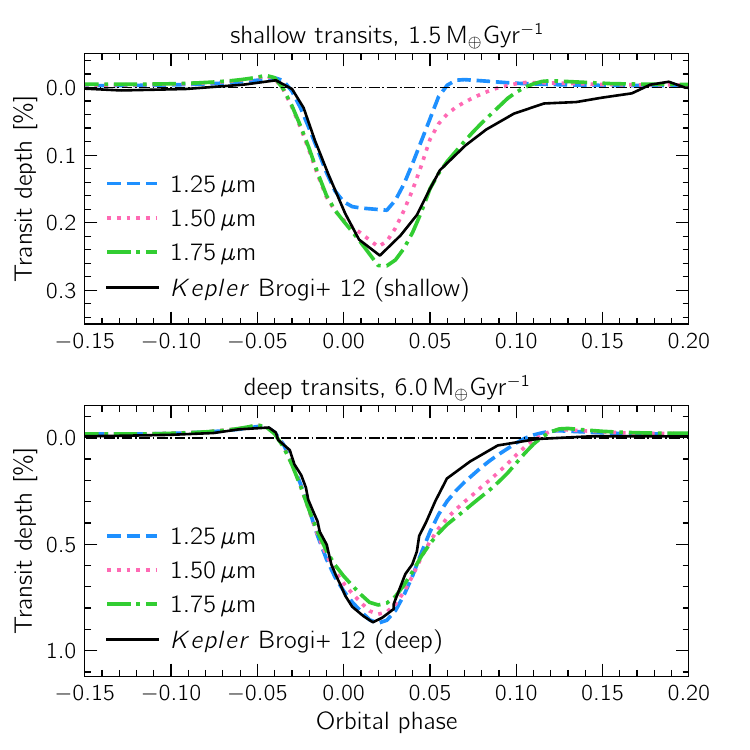}
\end{minipage}
\hspace{0.5cm}
\begin{minipage}[b]{0.48\linewidth}
\includegraphics[width=\linewidth]{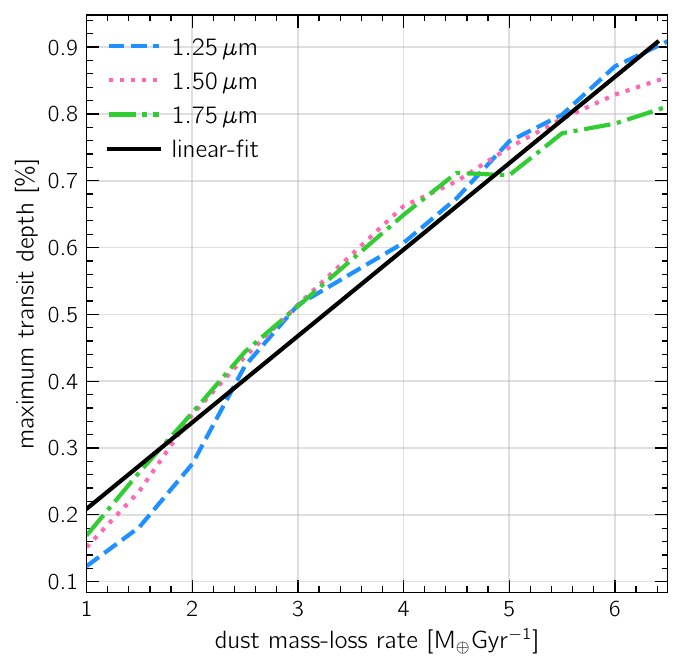}
\end{minipage}
 \caption{\bce{\textbf{Left}: Synthetic light curves for models at a constant planetary mass-loss rate of 1.5$\,M_{\oplus}\mathrm{Gyr}^{-1}$ ({\it top}) and 6.0$\,M_{\oplus}\mathrm{Gyr}^{-1}$ ({\it bottom}) at different initial dust grain sizes. The observed \textit{Kepler} light curve averaged over six quarters by \citet{Brogi2012} is shown as the black solid curve. Shallow transits correspond to observed light curves with transit depths between 0.2\,-\,0.5\,\%, deep transits correspond to observed light curves with transit depths larger than 0.8\,\% \citep[see,][]{Brogi2012}. \textbf{Right}: The maximum transit depth for the synthetic light curves of Mg$_{0.8}$Fe$_{1.2}$SiO$_4$ versus the dust mass-loss rate, at different initial dust grain sizes. The black solid curve represents a linear fit to the data.}}
 \label{fig:mdots}
\end{figure*}

\subsection{Corundum} 
\label{sec:corundum}
 
We find corundum dust grains with initial sizes below 2.0\,$\mu$m sublimate too fast to produce the observed light curve of KIC\,1255\,b. This is expected as small corundum grains achieve very high temperatures (Figure 4 in \citealt{Booth2023}), due to their low IR opacity. \bce{We find corundum could give origin to the observed light curve of KIC\,1255\,b with initial dust grain sizes of $\sim$3.5\,-\,5.5\,$\mu$m and an average mass-loss rate of 8.0\,$\,M_\oplus \mathrm{Gyr^{-1}}$ (Figure~\ref{fig:Al2O3})}. This result arises because as the particles get larger, the ratio between their optical and IR opacity falls resulting in lower temperatures and longer lifetimes; however, larger particles have a lower opacity and we must increase the mass-loss rate to match the observed transit depth. 

\bce{The initial dust grain sizes we estimate for dust composed of corundum are above the lower limit estimated from colour dependence measurements. The average dust mass-loss rate we estimate is within the findings of \citet{vanLieshout2016} but considerably higher than the theoretical models of \citet{PerezBecker2013} and \citet{Booth2023}. The estimate of the dust mass-loss rate in other studies \citep[e.g.][]{Kawahara2013}{}{} is generally obtained solely from the transit depth, ignoring any forward scattering effects. Forward scattering reduces the transit depth and therefore, higher mass-loss rates are needed to fit the observations (Appendix~\ref{ap:extvsscat}).}

\bce{As discussed by \citet{vanLieshout2016}, aluminium has a low cosmic abundance and is likely a minor component of rocky planets \citep[e.g.][]{Schaefer2009, Jura2014}. In addition to this, corundum is measured to have a relatively low abundance in the bulk silicate earth (BSE) composition (see Table 1.7 in \citealt{BSE}). Therefore, the existence of a corundum dust tail combined with such high mass-loss rates would be surprising. While corundum can fit the data, we consider it an unlikely candidate for the composition of the dust in the tail.}

\begin{figure}
\centering
\includegraphics[width=\linewidth]{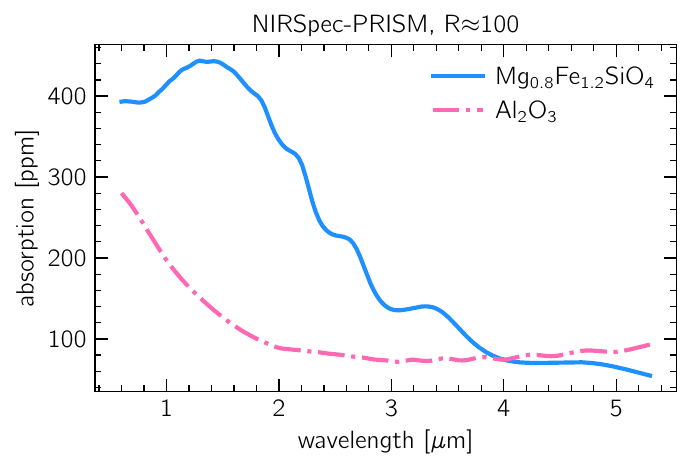}
\includegraphics[width=\linewidth]{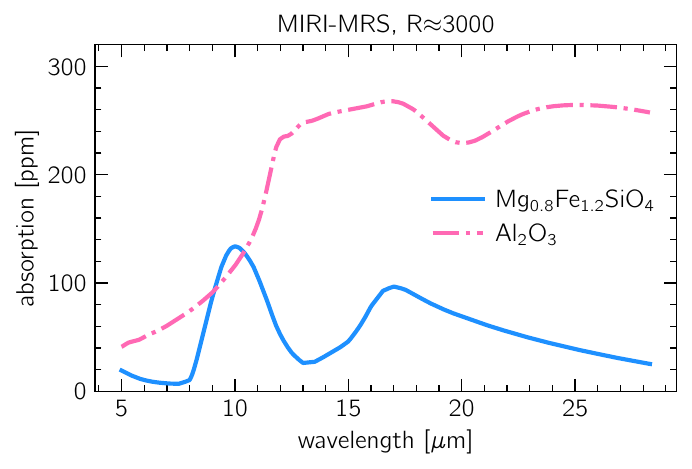}
\caption{\textbf{Top}: NIRSpec-PRISM synthetic absorption spectra of KIC\,1255\,b for the best-fit model of corundum (4.5$\,\mu$m and 8.0$\,M_\oplus \rm{Gyr}^{-1}$ - green dot-dashed curve in Figure~\ref{fig:Al2O3} left) and iron-rich olivine (1.25$\,\mu$m and 3.0$\,M_\oplus \rm{Gyr}^{-1}$ - pink dot-dashed curve in Figure~\ref{fig:Mg08Fe12SiO4} left). The resolution used was 100. \textbf{Bottom}: Same as the top panel but for the MIRI-MRS mode. The resolution used was 3000. Note the distinct absorption feature of the olivine at about 10\,$\mu$m.}
\label{fig:spectra}
\end{figure}

\subsection{Olivine and Pyroxene}
\label{sec:MgSi}
We have studied compounds we know are predominant in the Earth's mantle \citep[e.g.][]{BSE}: olivine, $\mathrm{(Mg,Fe)_2SiO_4}$[s], and pyroxene, $\mathrm{(Mg,Fe)SiO_3}$[s]. 

First, we consider the end-members of these two types of minerals. The iron end-member of olivine (fayalite), $\mathrm{Fe_2SiO_4}$\,[s], is too volatile in the entire parameter space studied, i.e. it sublimates too fast to explain an observable dusty tail. On the other hand, the magnesium end-members of olivine (forsterite), $\mathrm{Mg_2SiO_4}$[s], and pyroxene (enstatite), $\mathrm{MgSiO_3}$[s], have low optical to IR opacity ratio and do not achieve high enough temperatures to sublimate. The dust survives for multiple orbits, accumulating over time, and the transit profile never converges (see the right panel of Figure~\ref{fig:forsterite}). Additionally, the synthetic light curves produced show no agreement with the observations at any time-point (see the left panel of Figure~\ref{fig:forsterite}). The pre-transit brightening occurs earlier than expected for most of the parameter space. Furthermore, there is a significant post-transit brightening, which does not match the observations. Therefore, we rule out dust grains predominately composed of fayalite, forsterite or enstatite. 

We test other types of pyroxene (see Table~\ref{tab:dust}). For $\mathrm{Mg_{0.95}Fe_{0.05}SiO_3}$[s], the scenario is similar to that of forsterite and enstatite, i.e. the dust is too cold due to its low optical to IR opacity ratio. $\mathrm{Mg_{0.7}Fe_{0.3}SiO_3}$[s] and $\mathrm{Mg_{0.5}Fe_{0.5}SiO_3}$[s] behave similarly to fayalite - these compounds are too volatile and do not survive long enough to create the dusty-tail observed. Given the contrasting behaviours of the different types of pyroxene tested here, we believe low-iron pyroxene with an iron content between 0.05 and 0.3 should be explored in future work. 

The dust compositions tested which best fit with the observations of KIC\,1255\,b are magnesium-iron olivines, i.e. compounds of the form $(\text{Mg},\text{Fe})_{\,2}\text{SiO}_4$[s] with an iron content of at least 10\%. Figure~\ref{fig:Mg08Fe12SiO4} shows that for $\mathrm{Mg_{0.8}Fe_{1.2}SiO_4}$[s], the pre-transit brightening and the long-egress match the observation \bce{for dust grain sizes between 1.25$\,\mu$m and 1.75$\,\mu$m and an average planetary mass-loss rate of 3.0$\,M_{\oplus}\mathrm{Gyr}^{-1}$.} Similar results are obtained for Sri Lanka olivine, San Carlos olivine and $\mathrm{MgFeSiO_4}$[s]. 

These results are in agreement with the BSE abundances \citep[e.g][]{Kargel1993} and white dwarf measurements for evaporating exoplanets \citep[e.g][]{Bonsor2020}. \citet{Curry2023}  show that the observed dust is coming from a localised region where the planet has been evaporated down to. This supports the idea that the dust originates from the mantle of the planet, where magnesium-iron silicates like olivine and pyroxene are expected to be abundant. In addition to this, \citet{Bromley2023} argued iron-rich silicates are required to be able to produce an observable mass loss rate for catastrophically evaporating exoplanets.

\subsection{Estimating the average mass-loss rate}
\label{sec:mdots}
\bce{Since the transits are variable, and the precision on an individual transit is poor, we must compare our model to phase-folded, average light curves. Therefore, we must evaluate whether the mass-loss rate reported by our model, when compared to averaged light curves, is representative of the average mass-loss rate. We did this by computing the relationship between the mass-loss rates and the transit depth in our models. As shown in the right panel in Figure~\ref{fig:mdots}, we find the transit depth has an approximately linear relation with the dust mass-loss rate. A linear relationship implies that average light curves can be used to estimate average mass-loss rates. Additionally, we compared our synthetic light curves for Mg$_{0.8}$Fe$_{1.2}$SiO$_4$ to two sets of observed light curves of KIC\,1255\,b as divided by \citet{Brogi2012}: a set where the transit depths are shallow (0.2$\%$,-\,0.5$\%$) and a set where the transit depths are deep (>\,0.8\,$\%$). We find the best-fit for the shallow and deep transits require smaller and larger mass-loss rates, respectively, but the same range of dust grain sizes as for the average mass-loss rate case (left panel Figure~\ref{fig:mdots}). Therefore, our reported best-fit mass-loss rates to the average light curves are a reasonable approximation to the average mass-loss rate.}

\subsection{Synthetic JWST spectra: corundum vs magnesium-iron silicates}
\label{sec:jwst}
While we have argued corundum is an unlikely composition, we suggest that this can be directly confirmed with additional observations. Corundum and magnesium-iron silicates have distinct absorption features in the near-infrared and mid-infrared regions. In particular, silicates show a very broad absorption feature at about 10\,$\mu$m. Observing the dusty tails in these wavelengths can help us understand what the dust is composed of and validate our models. \bce{We produce synthetic JWST absorption spectra of KIC\,1255\,b for the $\mathrm{Mg_{0.8}Fe_{1.2}SiO_4}$[s] best-fit model with an initial dust grain size of 1.25$\,\mu$m and an average mass-loss rate of 3.0$\,M_\oplus \rm{Gyr}^{-1}$ (pink dot-dashed curve in Figure~\ref{fig:Mg08Fe12SiO4} left), and the corundum best-fit model with an initial dust grain size of 4.5$\,\mu$m and an average mass-loss rate of 8.0$\,M_\oplus \rm{Gyr}^{-1}$ (green dot-dashed curve in Figure~\ref{fig:Al2O3} left).} We use the wavelength range and resolution of the MIRI-MRS mode and the NIRSpec-PRISM mode. The spectra are shown in Figure~\ref{fig:spectra}. The synthetic spectra show that we would be able to distinguish if the dust is composed of corundum or magnesium-iron silicates with JWST observations if a bright target is identified \footnote{It is likely KIC 1255 b is too faint to perform this experiment with a small number of transits; however, K2-22 b may be bright enough.}. \rev{This inference is in line with the results of \citet{Bodman2018} and \citet{Okuya2020}, who also found JWST observations could help constrain the dust composition.}
\section{The leading dust tail of K2-22b and the outflow geometry}
\label{sec:K222b}
In addition to modelling KIC\,1255\,b, we also model K2-22\,b. As mentioned earlier, K2-22\,b shows a light curve with increases in the observed flux both before and after the transit \citep[][]{SanchisOjeda2015}. This agrees with a scenario where there is a tail of dust also leading the planet. K2-22\,b orbits an M-type star with $T_{\star}\approx3830 \, K$, $R_{\star}\approx0.58\,R_{\odot}$ and $M_{\star}\approx0.60\,M_{\odot}$, with an orbital period of approximately 9.2h \citep[][]{SanchisOjeda2015}{}{}.

\citet{SanchisOjeda2015} state a "substantial" ($\beta \gtrsim 0.05$) radiation pressure force necessarily blows the dust into orbits trailing the planet. However, they find if $\beta$ is sufficiently small ($\lesssim 0.02$) some dust can fall into faster orbits than that of the planet, creating a tail of dust leading the planet. 
\begin{figure*}
\centering
\begin{minipage}[b]{0.48\linewidth}
\includegraphics[width=\linewidth]{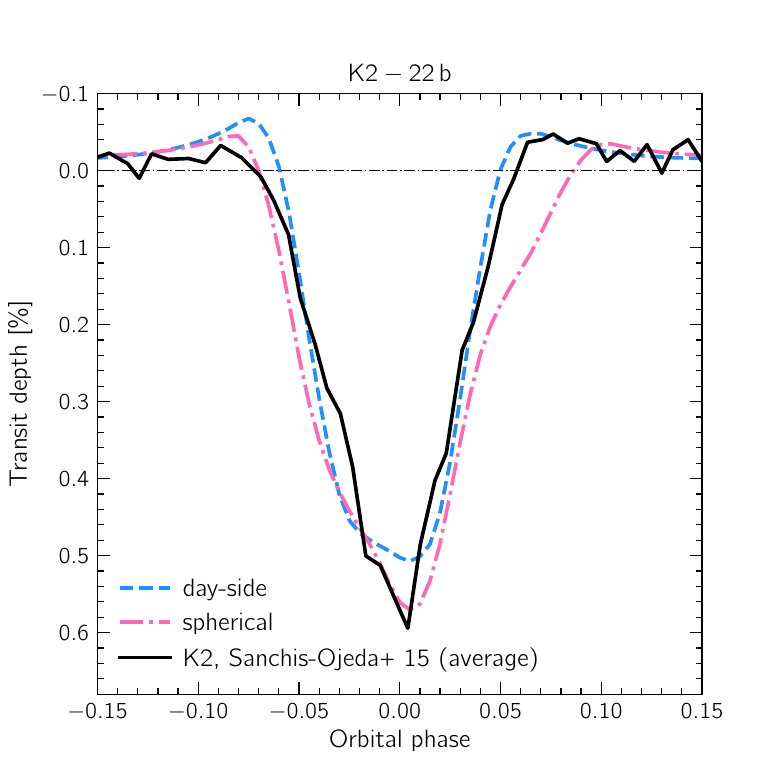}
\end{minipage}
\begin{minipage}[b]{0.48\linewidth}
    \includegraphics[width=\linewidth]{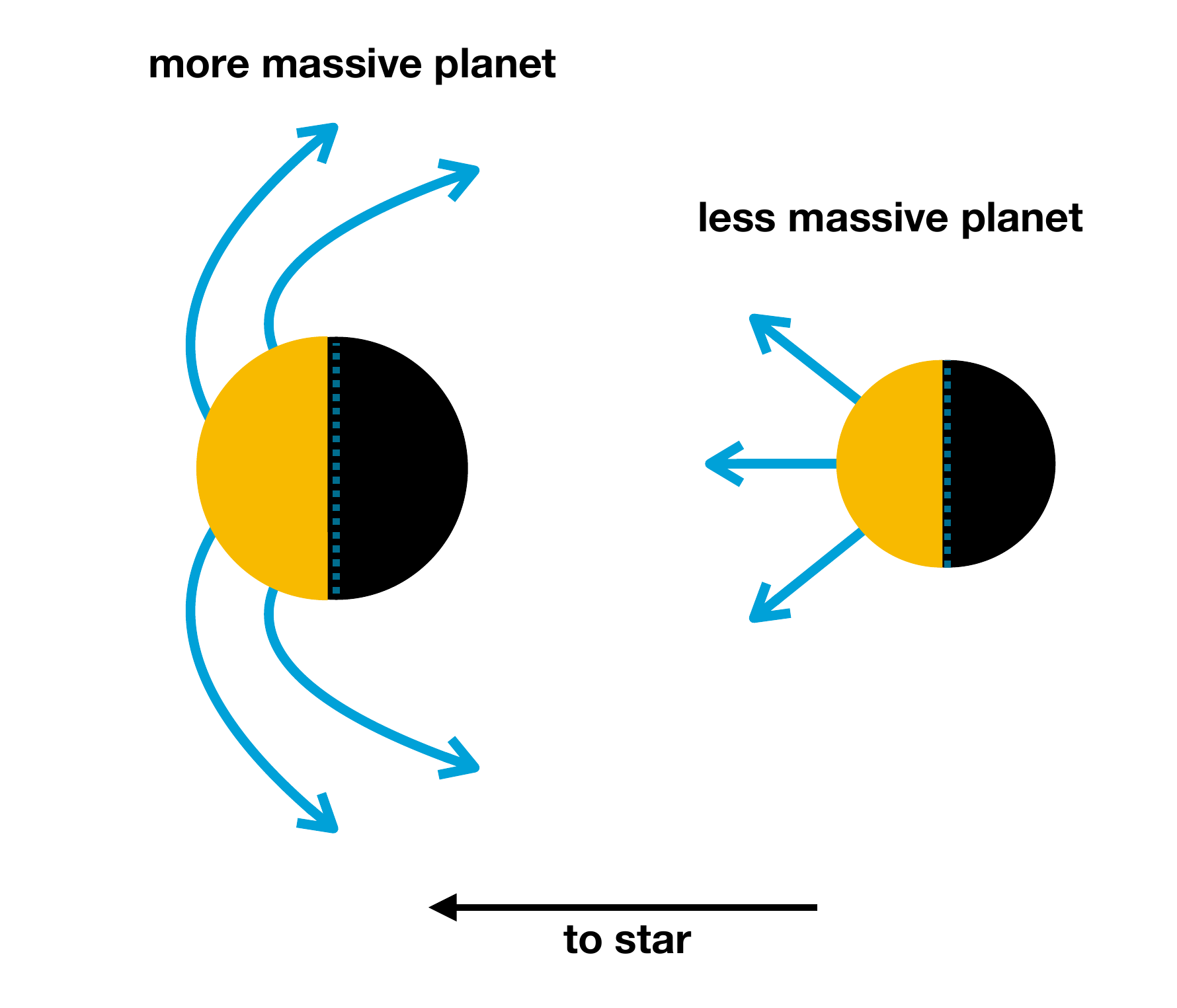}
\end{minipage}
\caption{\textbf{Left}: Synthetic light curves of K2-22\,b for dust composed of iron-rich olivine ($\mathrm{Mg_{0.8}Fe_{1.2}SiO_4}$[s]). The blue dashed curve is for a day-side outflow with an initial dust grain size of 1.5$\,\mu$m and a mass-loss rate of 3.0$\,M_\oplus \rm{Gyr}^{-1}$. The pink dot-dashed curve is for a spherical outflow with an initial dust grain size of 1.0$\,\mu$m and a mass-loss rate of 2.5$\,M_\oplus \rm{Gyr}^{-1}$. The observed \textit{K2} light curve is the black solid curve \citep[][]{SanchisOjeda2015}{}{}. \textbf{Right:} Cartoon of the dust outflow trajectory. The yellow represents the planet's day-side. The dust forms on the planet's day-side. If the planet is massive enough, the dust will follow a trajectory similar to that shown on the planet on the left - this is what the spherical outflow in our model depicts. If the planet is not massive enough, the dust will follow a trajectory similar to that shown on the planet on the right - this is what the day-side outflow in our model assumes.}
\label{fig:K222b}
\end{figure*}

\bce{The best-fit models we find for the K2-22\,b average light curve \citep[][]{SanchisOjeda2015}, assuming the dust is composed of $\mathrm{Mg_{0.8}Fe_{1.2}SiO_4}$[s], are for initial dust grain sizes $\sim$1.0\,-\,1.5$\mu$m and a mass-loss rate $\sim$2.5\,-\,3.0$\,M_\oplus \rm{Gyr}^{-1}$. These results compare favourably to \citet{Schlawin2021}, who observationally constrained the dust grain sizes of K2-22\,b to be larger than 0.5-1.0$\,\mu$m. Based on this measurement, they constrained the average mass-loss rate of K2-22\,b to be $\sim$ 1.6$\,M_\oplus \rm{Gyr}^{-1}$}.

\bce{Here, we investigate if the origin of the leading tail is a consequence of the launch geometry of the dusty outflow. We test two outflow geometries: a spherical outflow, where the super-particles leave from the Hill sphere radially outwards over the full $4\pi$, and a day-side outflow, where the super-particles leave the Hill sphere only from the day-side of the planet, radially outwards. As discussed in our methods, we always launch our super-particles from just outside the planet's Hill sphere. Physically, the outflow is always launched from the day-side surface of the planet; however, multi-dimensional simulations of photoevaporating planets with H/He-rich atmospheres \citep[e.g.][]{Stone2009,Owen2014,Tripathi2015} indicate the outflow can wrap around to the nightside, yielding a quasi-spherical outflow through the Hill sphere. Thus, we speculate a high-mass planet could have an outflow similar to those seen in these photoevaporation simulations, while a low-mass planet might have an outflow geometry that is day-side dominated at the Hill sphere (Figure~\ref{fig:K222b}, right).} 

\bce{We find the spherical outflow is a better fit for the pre-transit brightening of K2-22\,b. However, we find the day-side outflow to be a better fit for the egress and post-transit brightening (Figure~\ref{fig:K222b}, left). We find a post-transit brightening and a symmetric light curve (i.e. a leading tail of dust) can exist for both geometries in certain ranges of the parameter space, even though they don't fit the light curve perfectly. Thus, even in the spherical outflow scenario, this implies that changing the stellar parameters and the planet's orbital period is enough to give rise to a leading tail of dust. This result is a consequence of the strength of the radiation pressure force.  The radiation pressure is weaker for K2-22 compared to KIC\,1255 because K2-22 is cooler than KIC\,1255 by about 700K. Therefore, dust grains can occupy orbits closer to the star.
While we cannot produce a model light curve that matches all the features of K2-22b, we speculate that K2-22\,b's outflow geometry probably lies somewhere between a spherical and a day-side outflow. Coupling this launching geometry with a lower radiation pressure force to stellar gravitational force ratio than KIC\,1255\,b , would likely yield K2-22\,b's light curve.}

\bce {We find a day-side outflow at the level of the Hill sphere is not a good fit with the observations of KIC\,1255\,b (Appendix~\ref{ap:dayside}). }
Thus, the fact that KIC\,1255\,b is consistent with a spherical outflow and K2-22b with an outflow geometry that sits between our spherical and day-side models might imply KIC\,1255\,b is more massive than K2-22b. In this scenario KIC\,1255\,b is massive enough for a significant fraction of the escaping dust grains to wrap around the planet as they escape, as seen in hydrodynamic simulations \citep[e.g.][]{Stone2009,Owen2014,Tripathi2015}, while for K2-22\,b they are effectively unbound when they leave the planet's surface.

\begin{figure*}
\centering
\begin{minipage}[b]{\linewidth}
\includegraphics[width=\linewidth]{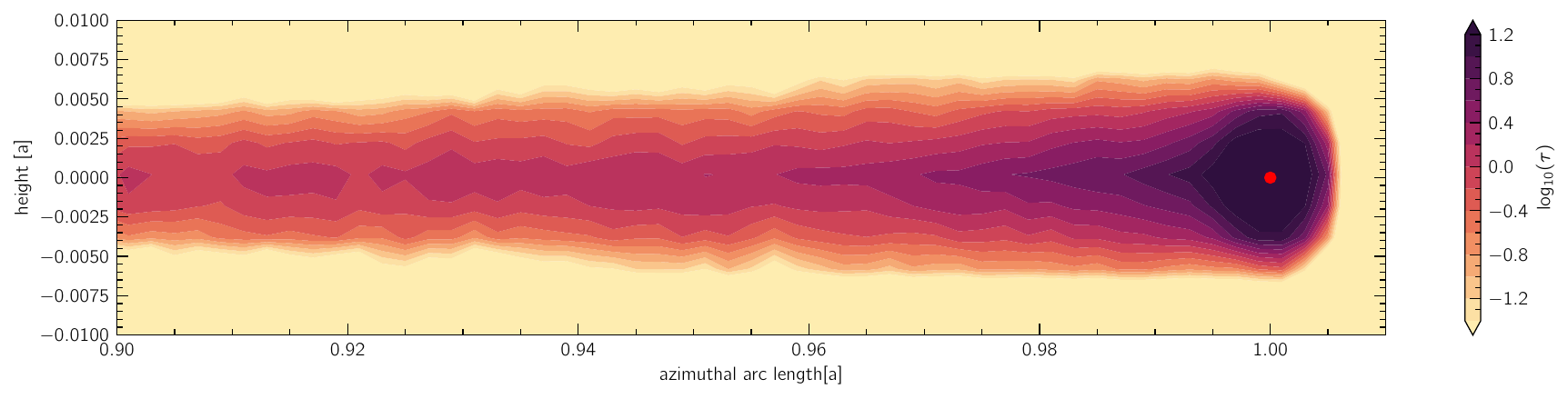}
\end{minipage}
 \caption{\bce{Optical depth density map at the radial edge of the optical depth grid ($R=1.10\,a$) for the best-fit $\mathrm{Mg_{0.8}Fe_{1.2}SiO_4}$[s] model (1.25$\,\mu$m, 3.0$\,M_\oplus \rm{Gyr}^{-1}$, dot-dashed pink curve in the left panel of Figure~\ref{fig:Mg08Fe12SiO4}). The red circle indicates the position of the planet. Note how the optical depth is moderate in the vicinity of the planet, implying attenuation of stellar light is important for controlling the dust's dynamics.}}
 \label{fig:tau_density_map}
\end{figure*}
\begin{figure}
\centering
\includegraphics[width=\linewidth]{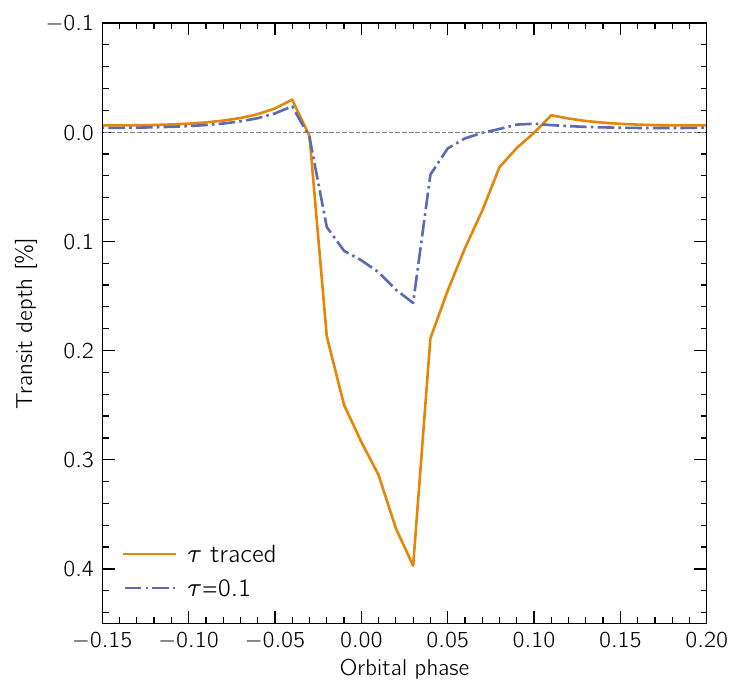}
 \caption{Comparison of two synthetic light curves for a model where the optical depth of the dust cloud is kept thin at 0.1 (dash-dotted curve) and a model where the optical depth is traced (solid curve).}
 \label{fig:tau_trace}
\end{figure}

\section{Optical depth and transit depth time variability}

As speculated by \citet{Rappaport2012}, \citet{PerezBecker2013} and \citet{vanLieshout2016}, we find the dust cloud to be optically thick in the vicinity of the planet (see Figure~\ref{fig:tau_density_map}). This affects the transit light curve of the planet and its tail in our models significantly, as shown in Figure~\ref{fig:tau_trace}. The transit depth for the models where the optical depth is traced is consistently larger than the transit depth in the models in which we assume a fixed low value for the optical depth. This is because the dust is cooler in the optically thicker environment, and the grains do not sublimate as fast. Although the optical depth also affects the radiation pressure force (i.e., $\beta$ parameter, see Equation~\ref{eq:beta}), its effect on the sublimation is stronger as the sublimation rate is exponentially dependent on the dust temperature (although radiation pressure is also exponentially sensitive to optical depth, our models only reach moderate optical depths).

Although we have included the optical depth evolution of the dusty tail, we do not find a transit depth time variability in our simulations. One of the limitations of our model is the fact that the planetary mass-loss rate is a free parameter and, therefore, independent of the optical depth of the dust cloud. In reality, the planetary mass-loss rate of the planet is dependent on the optical depth of the dust cloud. The rock vapour pressure is exponentially dependent on the surface temperature of the planet. So, a fractional change in the dust cloud's optical depth could cause large differences in the planetary mass-loss rate \citep[e.g.][]{PerezBecker2013}. \rev{Furthermore, we have disregarded the dynamical impact of the stellar wind on the dust grains as for micron-sized particles, the stellar wind pressure is generally negligible compared to the radiation pressure \citep[e.g.][]{Strubbe2006}. However, the stellar wind may become relevant for extremely large stellar wind rates. These rates can happen during, for example, coronal mass ejections, which can be a source of the observed variability \citep[cf.][]{Kawahara2013, Croll2015, Schlawin2018}.}

The observations of KIC\,1255\,b show that its transit depth variability presents no pattern \citep[e.g][]{vanLieshout2018}{}{}. Both \citet{PerezBecker2013} and \citet{Booth2023} have argued the variability might be driven by a cycle involving stellar insolation and mass loss. In addition to this, \citet{Booth2023} found that non-steady outflows can arise for fast dust growth rates and moderate optical depths. However, a chaotic behaviour, like the one the observations show, was not reproduced. Recently, \citet{Bromley2023} found a chaotic behaviour for the transit depth when they consider the planetary surface temperature to increase with increasing optical depth when the atmosphere is optically thin, and when it decreases with increasing optical depth when the atmosphere is optically thick. Such behaviour is only possible if, \bce{when the atmosphere is optically thin,} the dust has lower opacities in the visible wavelengths than in the infrared. This is because, in order to condense, the dust grains need to radiate away their energy more efficiently than they absorb visible light. In addition to this, for the planetary surface temperature to increase with increasing optical depth in an optically thin atmosphere, the dust needs to induce a greenhouse effect, which will raise the planetary surface temperature via infrared back-warming. Iron-poor silicate-rich dust presents the ideal visible-to-infrared opacity ratio to give rise to the conditions described above. When enough dust grains have condensed, and the atmosphere has therefore become optically thick to starlight, more iron can condense onto the grains. The dust will begin to absorb more efficiently in the visible than in the infrared, and the \bce{planetary surface temperature} will decrease, inducing what \cite{Bromley2023} denominate a nuclear winter. \citet{Bromley2023} state this cycle between the greenhouse effect and the nuclear winter is the most likely cause of the transit depth time variability.

\citet{Booth2023} discuss heterogeneous condensation might give rise to the transit depth time variability. They argue stable iron-free or iron-poor silicates will condensate first in the wind (e.g. $\mathrm{Mg_2SiO_4}$[s]) at low temperatures. This will allow iron to condense into these grains, causing the dust temperature to rise. However, iron evaporates more promptly than magnesium from silicates \citep[e.g.][]{Costa2017}{}{} which will increase the dust temperature, causing the iron content to decrease. This means the dust grains might reach a composition that is controlled by the feedback between the iron content and the temperature of the dust.

We find the dust is likely composed of magnesium-iron silicates (iron-poor or iron-rich) (Section~\ref{sec:dust_comp}). This is in agreement with both the greenhouse effect-nuclear winter cycle \citep{Bromley2023} and heterogeneous condensation \citep[][]{Booth2023}{}{}. A complete model which couples the dust formation to the outflow dynamics and tail morphology is needed to accurately study which processes are giving rise to the transit depth time variability. In addition to this, compositionally heterogeneous dust grains should be considered.

\label{sec:depth_variability}

\section{Summary}
\label{sec:summary}
Catastrophically evaporating rocky planets provide a unique opportunity to study the composition of rocky worlds. We have developed a self-consistent model of the dusty tails of catastrophically evaporating rocky planets. For the first time, we have introduced the optical depth evolution of the dust cloud in a model of this kind. We apply the model to two catastrophically evaporating exoplanets: KIC\,1255\,B and K2-22\,b. 

For both planets, we find the dust is likely composed of magnesium-iron silicates. \bce{The synthetic light curves of KIC\,1255\,b match the observed light curve for average (dusty) planetary mass-loss rates of $\sim 3$$\,M_{\oplus}\mathrm{Gyr}^{-1}$ and initial dust grain sizes between $\sim$ 1.25 and $\sim$ 1.75\,$\mu$m. Although corundum also produces models that fit the \textit{Kepler} observations, this is for very large initial dust grain sizes ($\sim$ 4\,$\mu$m) and large planetary mass-loss rates ($\sim$ 8$\,M_{\oplus}\mathrm{Gyr}^{-1}$). The existence of a corundum dust tail
combined with such high mass-loss rates would be surprising as aluminimum is likely a minor component of rocky planets \citep[e.g.][]{Schaefer2009}{}{}}. 
 JWST observations in the wavelength range of 1-28 $\mu$m could explicitly rule out corundum as the composition of the dust.

We show K2-22\,b probably has a different outflow geometry than KIC\,1255\,b. Thus, K2-22b is likely less massive than KIC\,1255\,b and presents a leading tail of dust. We find the leading tail of dust is likely a consequence of the outflow geometry and low radiation pressure force to stellar gravity force ratios ($\beta$). 

We find the dust cloud is marginally optically thick to stellar light in the vicinity of the planet, as first speculated by \citet{Rappaport2012}. This has a significant impact on the light curve, and therefore, future models of dusty tails should account for the optical depth of the tail, as was done in this work. Furthermore, \citet{Bromley2023} recently showed the observed transit depth time variability could have an origin in a greenhouse effect--nuclear winter cycle. Magnesium-iron silicates have the ideal visible-to-infrared opacity ratio to give rise to this cycle in the high mass-loss regime. In order to fully validate this hypothesis, we need to combine the model here developed with a model of the outflow dynamics and dust formation \citep[e.g.][]{Booth2023}{}{}.

\section*{Acknowledgements}
We thank the anonymous reviewer for their comments, which improved the manuscript.
We are grateful to Richard Booth, Subhanjoy Mohanty, Alfred Curry, Francisco Ard\'evol Mart\'inez, David A. Lewis  and Courtney J. Rundhaug for interesting discussions. We thank Eugene Chiang and Joshua Bromley for helpful comments on an earlier version of the manuscript. BCE is part of the CHAMELEON MC ITN EJD which received funding from the European Union`s Horizon 2020 research and innovation programme under the Marie Sklodowska-Curie grant agreement no. 860470. BCE was supported by a Royal Society 2020 Enhancement Award. JEO is supported by a Royal Society University Research Fellowship. MRJ acknowledges support from the European Union`s Horizon Europe programme under the Marie Sklodowska-Curie grant agreement No. 101064124, and funding provided by the Institute of Physics Belgrade, through the grant by the Ministry of Science, Technological Development, and Innovations of the Republic of Serbia. This project has received funding from the European Research Council (ERC) under the European Union`s Horizon 2020 research and innovation programme (Grant agreement No. 853022, PEVAP). For the purpose of open access, the authors have applied a Creative Commons Attribution (CC-BY) licence to any Author Accepted Manuscript version arising.

\section*{Data Availability}
The synthetic light curves data presented in the paper are available at the Electronic Research Data Archive (ERDA) of the University of Copenhagen at \href{https://sid.erda.dk/cgi-sid/ls.py?share_id=cceKwPeLFV}{https://sid.erda.dk/cgi-sid/ls.py?share\_id=cceKwPeLFV}. 




\bibliographystyle{mnras}
\bibliography{bib_paper}



\appendix

\section{\bce{KIC 1255 b, day-side outflow}}
\begin{figure*}
\centering
\begin{minipage}[b]{0.48\linewidth}
\includegraphics[width=\linewidth]{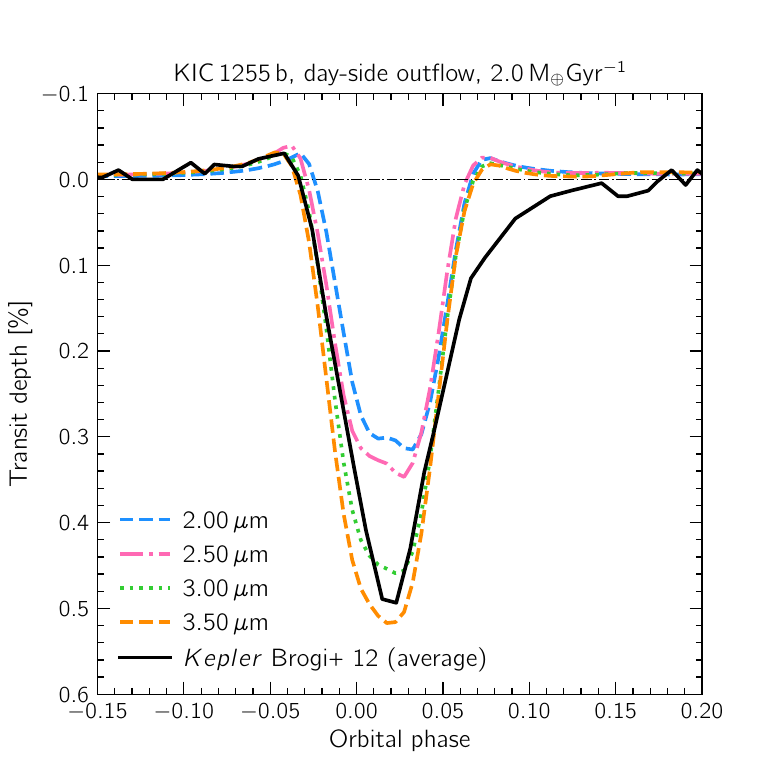}
\end{minipage}
\hspace{0.5cm}
\begin{minipage}[b]{0.48\linewidth}
\includegraphics[width=\linewidth]{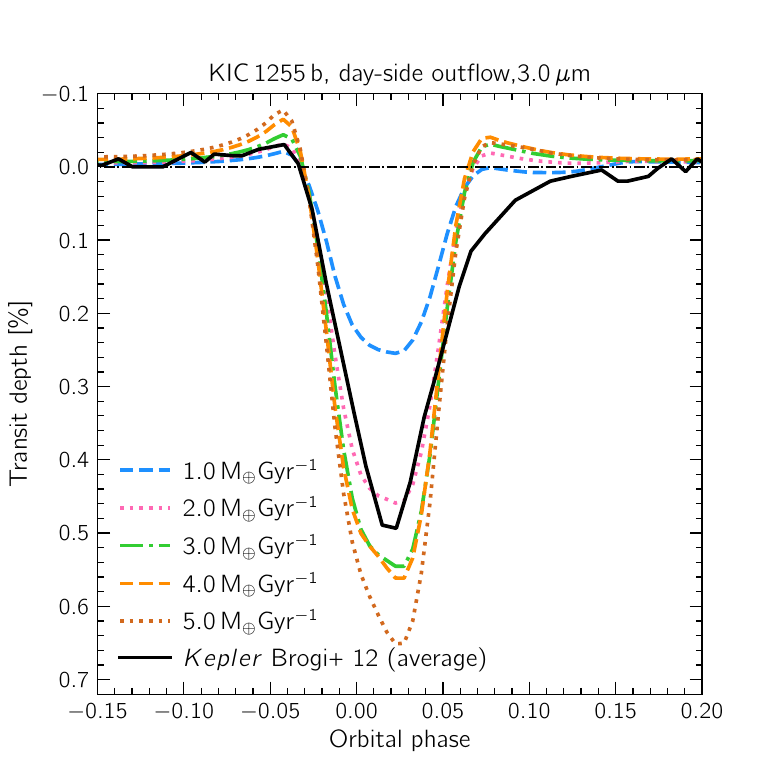}
\end{minipage}
 \caption{\bce{Synthetic light curves of KIC\,1255\,b for dust composed of iron-rich olivine ($\mathrm{Mg_{0.8}Fe_{1.2}SiO_4}$[s]) and a day-side dusty outflow. \textbf{Left}: Synthetic light curves for models at a constant planetary mass-loss rate of 2.0$\,M_{\oplus}\mathrm{Gyr}^{-1}$ and different initial dust grain sizes. The observed \textit{Kepler} light curve averaged over six quarters by \citet{Brogi2012} is shown as the black solid curve. \textbf{Right}: Synthetic light curves over 5 orbits at an initial dust grain size of 0.5$\,\mu$m and a planetary mass-loss rate of 1.0$\, M_{\oplus}\mathrm{Gyr}^{-1}$. }}
 \label{fig:KICdayside}
\end{figure*}
\label{ap:dayside}
\bce{The simulations for KIC$\,$1255$\,$b that assume a day-side outflow fail to reproduce the observed long-egress. This happens consistently in the parameter space explored. Figure~\ref{fig:KICdayside} shows the synthetic spectra for different dust mass-loss rates and initial dust grain sizes when assuming a day-side outflow for KIC$\,$1255$\,b$.}

\section{\bce{The extinction and forward scattering competition}}
\label{ap:extvsscat}
\begin{figure*}
\centering
\begin{minipage}[b]{0.48\linewidth}
\includegraphics[width=\linewidth]{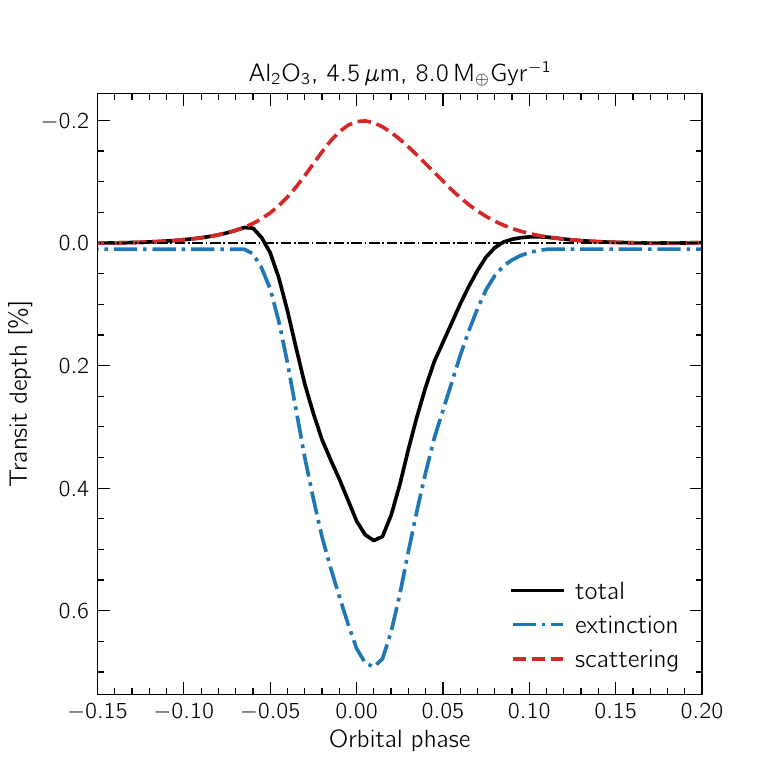}
\end{minipage}
\hspace{0.5cm}
\begin{minipage}[b]{0.48\linewidth}
\includegraphics[width=\linewidth]{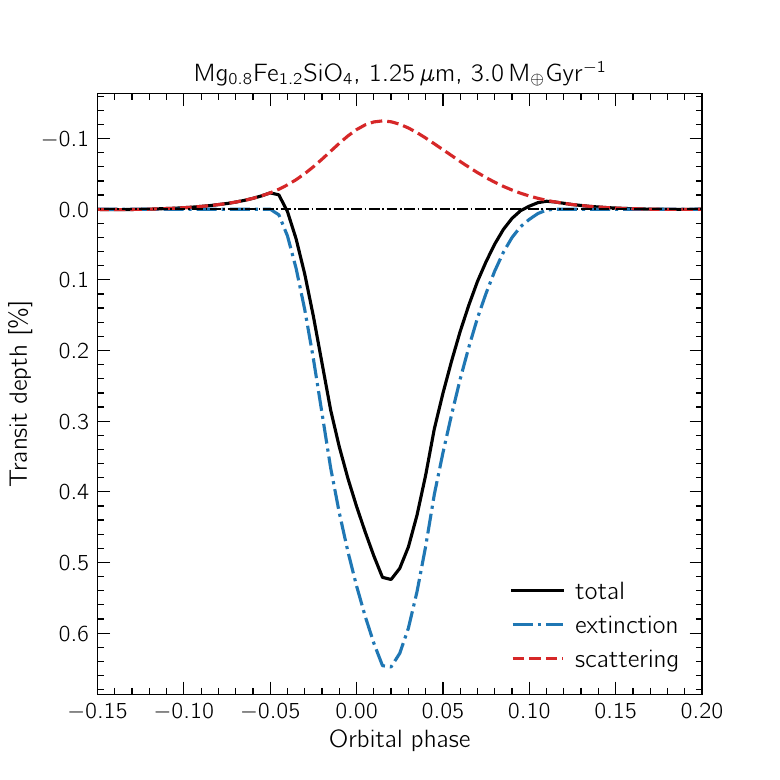}
\end{minipage}
 \caption{\bce{Synthetic light curves (black solid curves) of the best-fit corundum (\textit{left}) and iron-rich olivine (\textit{right}) models, and the scattering (dashed red curves) and extinction (dot-dashed blue curves) components. The components are smoothed to \textit{Kepler}'s long cadence. Note how the scattering component reduces the transit depth by a significant factor in both cases.}}
 \label{fig:fscat}
\end{figure*}

\bce{While the extinction increases the transit depth, forward scattering fills up the light curve. When including forward scattering in the modelling of dusty-tails, we require larger mass-loss rates to fit the observations (see also Figure~3, \citealt{vanLieshout2016}). Figure~\ref{fig:fscat} shows the contribution from scattering and extinction for the best-fit models of corundum and iron-rich olivine.}

\bsp	
\label{lastpage}
\end{document}